\documentclass[12pt,a4paper]{article}

\usepackage{amssymb}
\usepackage{epsf}

\begin{document}

\title{Friction forces on phase transition fronts}
\author{{\large Ariel M\'egevand\thanks{%
Member of CONICET, Argentina. E-mail address: megevand@mdp.edu.ar}} \\
{\normalsize \emph{IFIMAR (CONICET-UNMdP)}, }\\
{\normalsize \emph{Departamento de F\'{\i}sica, Facultad de Ciencias Exactas
y Naturales, UNMdP,} }\\
{\normalsize \emph{De\'an Funes 3350, (7600) Mar del Plata, Argentina} }}
\date{}
\maketitle

\begin{abstract}
In  cosmological first-order phase transitions, the microscopic
interaction of the phase transition fronts with non-equilibrium
plasma particles manifests itself macroscopically as friction forces.
In general, it is a nontrivial problem to compute these forces, and
only two limits have been studied, namely, that of very slow walls
and, more recently, ultra-relativistic walls which run away. In this
paper we consider ultra-relativistic velocities and show that
stationary solutions still exist when the parameters allow the
existence of runaway walls. Hence, we discuss the necessary and
sufficient conditions for the fronts to actually run away. We also
propose a phenomenological model for the friction, which interpolates
between the non-relativistic and ultra-relativistic values. Thus, the
friction depends on two friction coefficients which can be calculated
for specific models. We then study the velocity of phase transition
fronts as a function of the friction parameters, the thermodynamic
parameters, and the amount of supercooling.
\end{abstract}

\section{Introduction}

Cosmological phase transitions may have observable consequences. In
particular, first-order phase transitions provide a departure from
thermal equilibrium, which may give rise to a variety of cosmological
relics, such as the baryon asymmetry of the universe \cite{ckn93},
cosmic magnetic fields \cite{gr01}, topological defects \cite{vs94},
baryon inhomogeneities \cite{binhom,h95}, and gravitational waves
\cite{gw}. In a first-order phase transition, bubbles of the stable
phase nucleate and grow inside the supercooled phase. The velocity of
the phase transition fronts (bubble walls) is an important parameter
for the generation of cosmological relics. For instance, generating
sizeable gravitational waves requires high velocities, whereas the
generation of baryon number in the electroweak phase transition peaks
at small velocities.

The wall velocity is governed in principle by the pressure difference
between the two phases. However, the wall propagation disturbs the
plasma, and the latter resists the motion of the wall. This
opposition manifests itself in two ways. Microscopically, the
interactions of non-equilibrium plasma particles with the wall cause
a friction force on the latter \cite{fricthermal,js01,m00,bm09}.
Besides, the latent heat that is released at the phase transition
fronts causes temperature variations in the plasma
\cite{hidro,ikkl94,kl95,kl96}, which tend to diminish the pressure
difference between phases. These two mechanisms are called
microphysics and hydrodynamics, respectively.

Microphysics is a very difficult subject and for several years was
considered only in the non-relativistic (NR) limit. As a consequence,
phenomenological models have been used in order to extrapolate the
friction force to higher velocities. The general approach is based on
adding a covariant damping term $\eta u^{\mu}\partial_{\mu}\phi$ to
the field equation, where $u^{\mu}$ is the four-velocity of the
plasma (see, e.g., \cite{h95,ikkl94}). This gives a friction force of
the form $F_{\mathrm{fr}}\sim \eta \gamma_w v_w$, where $v_w$ is the
wall velocity and $\gamma_w=1/\sqrt{1-v_w^2}$. The parameter $\eta$
can be determined by considering the limit $v_w\to 0$ and comparing
with microphysics calculations.

Recently, the opposite limit was considered \cite{bm09}. The total
force acting on the electroweak bubble wall was derived for an
ultra-relativistic (UR) wall propagating with extremely large values
of the gamma factor. For such a fast moving wall the physics is quite
simpler than in the NR case, since the plasma is almost unaffected by
the passage of the wall. The resulting total force does not depend on
the wall velocity. As a consequence, if the force is positive, then
the wall runs away (i.e., as the wall propagates, $\gamma_w$ grows
linearly with the propagation distance). If the total force turns out
to be negative, then the initial assumption of an extremely
ultra-relativistic wall is incorrect and the wall must reach a
slower, stationary state. Thus, we have a simple criterion for the
existence of runaway solutions. The results of Ref. \cite{bm09}
motivated modifications to the usual phenomenological models for the
friction, so that the friction saturates for large $\gamma_w$
\cite{ekns10,hs13}. The modifications essentially amount to
eliminating the gamma factor from the friction force, so that
$F_{\mathrm{fr}}\sim\eta v_w$.

We wish to point out that the runaway condition found in Ref.
\cite{bm09} is only a necessary condition for the existence of the
runaway solution. It does not guarantee that the runaway solution
will actually be realized. Indeed, it is well known that different
propagation modes can exist for the same set of parameters (due to
nonlinear hydrodynamics). In particular, the runaway solution may
coexist with a detonation solution. Since detonations are presumably
stable, the wall probably will not run away whenever it can propagate
as a detonation. We also notice that a single-parameter
phenomenological model of the form $F_{\mathrm{fr}}=\eta v_w$ will
hardly give the correct values of the friction  force in both
ultra-relativistic and non-relativistic limits.

The main goals of this paper are the following. In the first place,
we wish to identify the friction force in the UR limit, in order to
treat very fast but stationary solutions.  This can be done by
decomposing the total force obtained in Ref. \cite{bm09} into a
driving force and a friction force. In principle, though, it is not
clear which is the correct decomposition, since there is no
velocity-dependent term in this force. Besides, hydrodynamics is
different in the stationary and runaway regimes. We wish to consider
the existence of stationary and runaway solutions and discuss the
necessary and sufficient conditions for the wall to run away.
Secondly, we wish to construct a well-motivated phenomenological
model, which interpolates between the NR and UR limits. The model
will have two free parameters and will allow us to treat stationary
solutions with intermediate velocities. We  also wish to discuss the
case of phase transitions with a large amount of supercooling, which
favor fast-moving phase transition fronts.

The plan is the following. In section \ref{hydro} we briefly review
the dynamics of cosmological phase transitions and the hydrodynamic
solutions for stationary fronts. In section \ref{micro} we discuss
the friction forces. We introduce an ultra-relativistic friction
parameter and we discuss the conditions for the wall to run away. In
section \ref{runaway} we discuss a phenomenological model for the
friction. In section \ref{result} we consider the bag equation of
state to obtain analytical formulas for the wall velocity and for the
runaway conditions. We also analyze the general dependence of the
bubble wall velocity on the friction and thermodynamic parameters.
Finally, in section \ref{strong} we discuss the case of strong
supercooling. We summarize our results in section \ref{conclu}.

\section{Phase transition dynamics} \label{hydro}

A cosmological phase transition is described in general by a scalar
field $\phi$ which acts as an order parameter. We shall consider the
case in which $\phi$ is a Higgs field. The finite temperature
behavior of the system is determined by the free energy density
(finite-temperature effective potential)
\begin{equation}
\mathcal{F}(\phi,T)=V(\phi) +V_T(\phi),  \label{ftot}
\end{equation}%
where $V(\phi)$ is the zero-temperature effective potential and
$V_T(\phi)$ the finite-temperature correction. To one-loop order, the
latter is given by \cite{quiros}
\begin{equation}
V_T(\phi) =\sum_{i}\pm {g_{i}}
\int_{0}^{\infty }\frac{d^3p}{(2\pi)^3}\log \left( 1\mp e^{-E_i/T} \right).
\label{f1loop}
\end{equation}
where the sum runs over particle species, $g_{i}$ is the number of
degrees of freedom of species $i$, the upper sign stands for bosons,
the lower sign stands for fermions, and $E_i=\sqrt{p^2+m_i^2(\phi)}$.
Here, $m_i$ are the field-dependent masses. There is also a
correction from the resummed daisy diagrams for bosons. For our
discussions it is not necessary to consider the exact form of this
correction. Its effect is to modify the cubic term in the expansion
of $V_T$ in powers of $m_i/T$, as we shall comment below.

In thermal equilibrium, the field lies at a minimum of the free
energy. If there are several minima, then different phases are
possible for the system, and phase transitions may occur. In the
simplest case we have  a high-temperature minimum $\phi_+(T)$ (in
general, $\phi_+\equiv 0$) and a low-temperature one $\phi_-(T)$. For
a first-order phase transition, these two minima coexist in a certain
range of temperatures, separated by a barrier. All the properties of
a given phase are derived from $\mathcal{F}$, once it is evaluated at
a minimum. Thus, the two phases are characterized by the free energy
densities
\begin{equation}
\mathcal{F}_{+}(T)\equiv \mathcal{F}(\phi_+(T),T),\;\;
\mathcal{F}_{-}(T)\equiv \mathcal{F}(\phi
_{-}(T),T),
\end{equation}
which give different equations of state (EOS). The energy density is
given by $\rho_{\pm} \left( T\right) =
\mathcal{F}_{\pm}(T)-T\mathcal{F}_{\pm}^{\prime }(T)$, where a prime
indicates a derivative with respect to $T$. The pressure is given by
$p_{\pm}(T)=-\mathcal{F}_{\pm}(T)$. The enthalpy density is given by
$w_{\pm}=\rho_{\pm}+p_{\pm}$, and the entropy density by
$s_{\pm}=w_{\pm}/T$. The speed of sound is given by $ c_{\pm}^{2}(T)
=\partial p_{\pm}/\partial \rho_{\pm}=p_{\pm}^{\prime}(T)/\rho_{\pm}
'(T)$. The phases are in equilibrium at the critical temperature
$T_c$, defined by $\mathcal{F}_{+}(T_{c})=\mathcal{F}_{-}(T_{c})$.
The energy density difference at $T=T_c$ is called the latent heat
$L\equiv\rho _{+}\left( T_{c}\right) -\rho _{-}\left( T_{c}\right)$.

The phase transition occurs, in principle, when the Universe reaches the
critical temperature. However, the nucleation rate vanishes at $T=T_c$.
Bubbles of the stable phase effectively begin to nucleate at a temperature
$T_n$ below $T_c$ \cite{c77}. The nucleated bubbles expand due to the
pressure difference between the two phases (at $T<T_c$). The phase
transition fronts move with a velocity $v_w$ which depends on the pressure
difference and the friction with the surrounding plasma. The release of
latent heat causes local reheating and bulk motions of the plasma.
Considering the development of a phase transition involves solving a set of
coupled equations for several quantities, each of which is not easy to
compute, such as the nucleation rate, the wall velocity, and the
temperature. The latter varies due to the expansion of the universe and the
release of energy at the phase transition (for the dynamics of phase
transitions see, e.g.,  \cite{ptdynamics}). Here we shall concentrate on the
motion of phase transition fronts for a given nucleation temperature $T_n$.

Thus, the relevant variables are the scalar field
$\phi(\mathbf{x},t)$, the temperature $T\left( \mathbf{x},t\right)$,
and the velocity $v\left( \mathbf{x},t\right) $ of the plasma. We
shall consider the thin-wall approximation for the field profile.
Thus, we shall assume that the field varies in an infinitely thin
region, outside which  $\phi$ is a constant. For the macroscopic
treatment this is a good approximation, since the wall width is much
smaller than the width of the fluid profiles \cite{ms12}. We shall
consider planar-symmetry fronts moving in the $z$ direction (see
\cite{lm11} for a discussion on considering different wall
geometries). As a consequence of the friction with the plasma, the
bubble wall often reaches a terminal velocity in a very short time
after bubble nucleation. In the rest of this section we shall review
the hydrodynamics for stationary phase transition fronts.

\subsection{Hydrodynamics for stationary fronts}

The equations for the fluid variables can be obtained from the conservation
of the energy-momentum tensor, $\partial _{\mu }T^{\mu \nu }=0$. For
stationary profiles moving at constant velocity, it is useful to consider
the rest frame of the front. The relation between the fluid variables on
each side of the wall is well known \cite{landau},
\begin{eqnarray}
w_{-}v_{-}\gamma _{-}^{2} &=&w_{+}v_{+}\gamma _{+}^{2},  \label{disc1} \\
w_{-}v_{-}^{2}\gamma _{-}^{2}+p_{-} &=&w_{+}v_{+}^{2}\gamma _{+}^{2}+p_{+},
\label{disc2}
\end{eqnarray}%
where $\gamma_{\pm}=1/\sqrt{1-v_{\pm}^2}$, and we have used a $+$
sign for variables  in front of the wall and a $-$ sign for variables
behind the wall (which correspond to the $+$ and $-$ phases,
respectively). For an infinitely thin interface, Eqs.
(\ref{disc1}-\ref{disc2}) determine the discontinuity of the fluid
profiles at the phase transition front. There can also be
discontinuities in the fluid profiles away from the bubble wall.
These are called \emph{shock fronts}. In such surfaces, the
temperature and fluid velocity are discontinuous but the EOS is the
same on both sides. In the reference frame of shock fronts Eqs.
(\ref{disc1}-\ref{disc2}) still apply.

Since there is no characteristic distance scale in the fluid
equations, it is usual to assume the \emph{similarity condition}
\cite{landau}, namely, that $w,p$ and $v$ depend only on $\xi =z/t$.
In the planar-symmetry case we have either constant solutions $ v(
\xi ) =\mathrm{constant}$, or a  ``rarefaction wave'' solution
$v_{\mathrm{rar}}( \xi ) =({\xi -c})/({1-\xi c})$, where $c$ is the
speed of sound (see, e.g., \cite{lm11,ms12} for recent discussions).
The fluid velocity profile is constructed from these solutions, using
the matching conditions (\ref{disc1}-\ref{disc2}) and appropriate
boundary conditions.

For a given set of thermodynamical parameters, Eqs. (\ref{disc1}) and
(\ref{disc2}) give $v_{+}$ as a function of $v_{-}$. The solutions have two
branches, called \emph{detonations }and \emph{deflagrations} (see, e.g.,
\cite{s82}). For detonations the incoming flow is faster than the outgoing
flow ($|v_{+}|>|v_{-}|$) and is supersonic ($|v_{+}|>c_+$) for any value of
$v_-$ in the range $ -1<v_{-}<0$. For deflagrations, we have
$|v_{+}|<|v_{-}|$ and $|v_{+}|<c_+$. At $|v_{-}|=c_{-}$, the value of
$|v_+|$ is a minimum for detonations and a maximum for deflagrations. At
this point the hydrodynamic process is called a \emph{Jouguet} detonation or
deflagration. We denote the corresponding values by
$|v_{+}|=v_{J}^{\mathrm{\det}}$ and $|v_{+}|=v_{J}^{\mathrm{def}}$,
respectively. The hydrodynamic process is called \emph{weak} if the
velocities $v_{+}$ and $v_{-}$ are either both supersonic or both subsonic.
Otherwise, the hydrodynamic process is called \emph{strong}.

The fluid temperature and velocity profiles depend on the boundary
conditions far behind the wall (at the center of the bubble) and far
in front of the wall, where information on the bubble has not arrived
yet. Thus, the fluid velocity vanishes far behind and far in front of
the wall. Furthermore, the temperature far in front of the wall is
given by the temperature $T_{n}$ at which the bubble nucleated. Three
kinds of solutions are compatible with these requirements: a weak
detonation, for which the wall is supersonic, a Jouguet deflagration,
which is also supersonic, and a weak deflagration, which is subsonic.

The relevant properties of these solutions are the following (see,
e.g., \cite{ms12,lm11} for details). For the supersonic detonation,
the fluid in front of the wall is unperturbed and we have
$v_{w}=-v_+, T_+=T_n$. Hence, we have $v_w\geq v_J^{\mathrm{det}}$.
The fluid profile behind the wall is given by the rarefaction wave.
The strong detonation solution must be discarded because it is not
possible to construct a fluid profile for it. For the deflagration,
the fluid velocity in front of the wall is not at rest, and there is
a supersonic shock front preceding the wall. Beyond the shock, the
fluid is still unperturbed at temperature $T_n$. The solution between
the phase-transition and shock fronts is a constant. The boundary
condition for deflagrations is often considered to be that the fluid
behind the wall is at rest. In this case, we have $v_{w}=-v_{-}$.
This ``traditional'' deflagration can be either weak (and subsonic),
or strong (and supersonic). The limit between these solutions is a
Jouguet deflagration with $v_w=c_-$. It has been argued that strong
deflagrations are unstable\footnote{Even weak deflagrations may be
unstable \cite{unstable}.} and should not be considered
\cite{ikkl94,kl95,kl96}. However, supersonic deflagrations that are
not strong can exist \cite{kl95} if we relax the boundary condition
that the fluid is at rest behind the wall. Thus, if the condition
$v_{-}=-v_{w}$ is replaced by $v_-=-c_-$, we have a Jouguet
deflagration with $|v_{+}|=v_{J}^{\mathrm{def}}<c_{+}$. In this case
the wall is followed by a rarefaction wave. Since the wall moves at
the speed of sound with respect to the fluid behind it, and the fluid
also moves with respect to the center of the bubble, the wall
velocity is supersonic. This solution fills the velocity gap between
the weak deflagration and the detonation, $c_-\leq v_w\leq
v_J^{\mathrm{det}}$.

\section{Microphysics} \label{micro}

The forces acting on the bubble wall can be derived from the equation
of motion for the background field $\phi$. It is usual to consider
the WKB approximation, which makes sense if the scale of variation of
$\phi$ is not too short. We have \cite{fricthermal}
\begin{equation}
\partial _{\mu }\partial ^{\mu }\phi +\frac{\partial V}{\partial
\phi }+\sum_i g_i\frac{dm^2_i}{d\phi}\int\frac{d^3p}{(2\pi)^32E_i}f_i(p,x)=0,
\label{eqmicro}
\end{equation}
where $f_i$ is the distribution function of particle species $i$. In
general, a distribution function can be written as
\begin{equation}
f_i(p)=f_i^{\mathrm{eq}}(p)+\delta f_i(p,x), \label{decomp}
\end{equation}
where $f_i^{\mathrm{eq}}(p)=1/(e^{E_i/T}\mp 1)$ is the equilibrium
distribution function and $\delta f_i$ is  a deviation. Inserting this
decomposition into Eq. (\ref{eqmicro}), the term $f_i^{\mathrm{eq}}$ just
gives $\partial V_T/\partial\phi$ (i.e., the finite-temperature corrections
to the effective potential) and we obtain
\begin{equation}
\partial _{\mu }\partial ^{\mu }\phi +\frac{\partial \mathcal{F}}{\partial
\phi }+\sum_i g_i\frac{dm^2_i}{d\phi}\int\frac{d^3p}{(2\pi)^32E_i}\delta f_i=0.
\label{eqmicro2}
\end{equation}
We shall assume for simplicity that the phase transition occurs from a
vanishing minimum to a non-vanishing one, as in the usual symmetry braking
case. We shall denote $\phi_0$ the non-vanishing minimum, i.e.,
\begin{equation}
\phi_+(T)\equiv 0, \;\; \phi_-(T)\equiv \phi_0(T)
\end{equation}

An equation for the bubble wall can be obtained if we assume a fixed
field profile centered at the wall position $z_w(t)$. Thus, consider
the dependence $\phi(z,t)=\phi [\gamma_w(z-z_w)]$. The exact form of
the one-variable function $\phi(z)$ is not relevant. It is often
assumed to be given by a $\tanh$. It varies from the value
$\phi(-\infty)=\phi_0$ (inside the bubble) to $\phi(+\infty)=0$
(outside the bubble). The variation occurs in a small region of width
$l_w$ (the bubble wall). In the reference frame which instantaneously
moves with the wall, in which $z_w=0$ and $\dot{z}_w=0$, we have
$\gamma_w=1,\dot{\gamma}_w=0$ and $\ddot{\gamma}_w=\ddot{z}_w^2$,
where a dot indicates a  derivative with respect to $t$. The first
term in Eq. (\ref{eqmicro2}) gives $(\ddot{z}_w^2
z-\ddot{z}_w)\phi'(z)-\phi''(z) $, where a prime indicates a total
derivative with respect to $z$. Notice that the function $\phi'(z)$
has a peak inside the wall and vanishes outside. We can define the
wall position so that $\int z\phi'^2 dz=0$. Then, if we multiply Eq.
(\ref{eqmicro2}) by $\phi'$ and integrate across the wall, we obtain
\begin{equation}
\sigma \ddot{z}_w=F_{\mathrm{dr}}/A + F_{\mathrm{fr}}/A, \label{fuerzas}
\end{equation}
where $\sigma\equiv\int \phi'^2 dz$ is the surface tension, and
\begin{eqnarray}
\frac{F_{\mathrm{dr}}}{A} &=& \int \frac{\partial
\mathcal{F}(\phi,T)}{\partial\phi}\frac{d\phi}{dz}dz, \label{driving}
\\
\frac{F_{\mathrm{fr}}}{A}&=&\sum_i g_i\int dz\frac{dm^2_i}{dz}
\int\frac{d^3p}{(2\pi)^32E_i}
\delta f_i \label{fricgral}
\end{eqnarray}%
are the forces per unit area acting on the wall.

The force $F_{\mathrm{fr}}$ depends on the interactions of the plasma
particles with the wall and may be regarded as a friction force.
Indeed, for small wall velocity $v_w$ (in the reference frame of the
bubble center), this force turns out to be proportional to $v_w$. As
a consequence, the wall may reach a terminal velocity. The force
$F_{\mathrm{dr}}$ does not depend on microphysics details, and may be
regarded as the driving force.

\subsection{The driving force} \label{drf}

To see the behavior of this force, let us consider a wall which moves
very slowly, so that equilibrium can be assumed  and
$F_{\mathrm{fr}}$ vanishes in Eq. (\ref{fuerzas}). For such a slow
wall, the temperature will be homogeneous, $T(z)=$ constant. Hence,
the integral in Eq. (\ref{driving}) yields
$\mathcal{F}(\phi_+,T)-\mathcal{F}(\phi_-,T)$. Thus, we obtain the
equation $\sigma \ddot{z}_w=p_-(T)-p_+(T)$, which is positive for
$T<T_c$. Therefore, if the wall is initially moving very slowly, it
will accelerate due to the pressure difference. As a consequence, the
wall motion will cause departures from local equilibrium and  a
friction force will appear. Besides, inhomogeneous reheating will
arise, which affects the driving force.

For the integral in Eq. (\ref{driving}) we may use the identity
$({\partial \mathcal{F}}/{\partial \phi }) ({d\phi }/{dz})=
{d\mathcal{F}}/{dz}-({\partial \mathcal{F}}/{\partial T})({dT}/{dz})$
\cite{ikkl94,ms09}, so that the driving force can be expressed as
\begin{equation}
\frac{F_{\mathrm{dr}}}{A}=p_{-}(T_-)-p_{+}(T_+)+\int_{T_-}^{T_+} s(\phi,T)
dT, \label{dr}
\end{equation}%
where the entropy density is given by $s=-{\partial
\mathcal{F}}/{\partial T}$. In order to obtain analytical results, in
this paper we shall use approximations for integrals across the wall.
Our aim is to obtain expressions which only depend on variables
defined outside the wall, namely, the values of $v_{\pm}$, $T_{\pm}$,
etc., which can be obtained using Eqs. (\ref{disc1}-\ref{disc2}), the
EOS, and appropriate boundary conditions. For the integral in Eq.
(\ref{dr}), we shall approximate the entropy density by the average
value $\langle s\rangle=(s_{+}+s_{-})/2$. This gives the
approximation \cite{ms09}
\begin{equation}
\frac{F_{\mathrm{dr}}}{A}=p_{-}(T_-)-p_{+}(T_+)+
\langle s\rangle\left( T_{+}-T_{-}\right)
.  \label{draprox}
\end{equation}%
If we neglect hydrodynamics and consider a homogeneous temperature,
we obtain the pressure difference $p_{-}(T)-p_{+}(T)$. This is very
sensitive to the departure of $T$ from $T_c$. Besides,
$p_{-}(T_-)-p_{+}(T_+)$  is very sensitive to the difference
$T_+-T_-$. Indeed, the pressure difference may be positive or
negative depending on the value of $T_+-T_-$. The general effect of
hydrodynamics is to slow down the wall \cite{ms09,kn11}.

\subsection{Friction force}

The friction force $F_{\mathrm{fr}}$ is in general much more
difficult to calculate than $F_{\mathrm{dr}}$. It has been
extensively studied for small wall velocities
\cite{fricthermal,js01}, since temperature gradients can be neglected
and the deviations from equilibrium can be assumed to be small. Even
in this case, computing the deviations $\delta f_i$ involves solving
a complex system of Boltzmann equations. Recently \cite{bm09}, the
total force acting on the  bubble wall was derived for a wall which
propagates ultra-relativistically. The treatment turns out to be
quite simpler.

\subsubsection{Non-relativistic limit}

For small velocities we have $\delta f_i \propto v_w$, and Eq.
(\ref{fricgral}) gives a friction term of the form
\begin{equation}
\frac{F_{\mathrm{fr}}}{A}=-\eta_\mathrm{NR} v_w.
\label{fricnr}
\end{equation}
The non-relativistic friction coefficient $\eta_\mathrm{NR}$ depends
on the couplings of the particles to the Higgs, and also on the
interactions of plasma particles, which tend to restore the
equilibrium. Effective interaction rates $\Gamma_{ij}\sim T$ appear
in the equations for $\delta f_i$, coming from the collision integral
in the Boltzmann equation. The ``thick wall'' limit $\Gamma\gg 1/l_w$
is often used, which leads to analytical results (see, e.g.,
\cite{js01}). Here we shall use the results from the simplified
treatment of Refs. \cite{m04,ms10}. For masses of the form
$m_i^2=h_i^2\phi^2+\mu_i^2$ we have
\begin{equation}
\eta_\mathrm{NR} \approx\sum_{i}\frac{g_{i}h_{i}^{4}}{\Gamma }\int_{-\infty }^{+\infty
}c_{1i}^{2}(\phi)\, \phi ^{2}\phi ^{\prime 2}dz.  \label{eta}
\end{equation}%
where $\Gamma$ is an average interaction rate (typically,
$\Gamma\lesssim 10^{-1}T$), and the function $c_1(\phi)$ is given by
\begin{equation}
c_{1}\equiv \frac{1}{T^{2}}\int \frac{d^{3}p}{\left( 2\pi \right) ^{3}E}
\frac{e^{E/T}}{(e^{E/T}\mp 1)^2}.
\end{equation}%
Notice the strong dependence of the coefficient $c_1$ on the ratio
$m/T$. Different limiting cases have been analyzed in Ref.
\cite{ms10}. For a ``typical'' phase transition with $\phi\sim T\sim
T_c$, one can often use an expansion in powers of $m/T$. To lowest
order, we have $c_1=\log \chi /2\pi ^{2} $, where $\chi =2$ for
fermions and $\chi =T/m$ for bosons. Hence, in this case the
dependence on $\phi$ is at most logarithmic, and we can regard $c_1$
as a constant for the integral in (\ref{eta}). We obtain
\begin{equation}
\eta_\mathrm{NR}
= \frac{\hat{\eta}_\mathrm{NR}}{T} \int \phi ^{2}\phi ^{\prime 2}dz,
\label{etatth}
\end{equation}
where
\begin{equation}
\hat{\eta}_\mathrm{NR}\equiv\sum_i \frac{g_{i}h_{i}^{4}}{\Gamma/T }
\left( \frac{\log \chi
_{i}}{2\pi ^{2}}\right) ^{2} \label{etahatnr}
\end{equation}
The dimensionless coefficient $\hat{\eta}_\mathrm{NR}$ depends on
details of the model, whereas the integral in Eq. (\ref{etatth})
depends only on the wall shape. This integral can be estimated as
$\int \phi ^{2}\phi ^{\prime 2}dz\sim\phi_0 ^{2}\sigma \sim
\phi_0^4/l_w$, where $l_w$ is the wall width. Typically, $l_w\gtrsim
T^{-1}$. These rough approximations give in principle the correct
parametric behavior, but omit numerical factors which may be as high
as $\sim 10$. To obtain quantitatively useful values, we may
``calibrate'' the friction with a known model \cite{hs13,hs12}. In
our approximation (\ref{etahatnr}), this can be accomplished by
choosing a suitable value of $\Gamma/T$. For instance, considering
the top, Z and W contributions to (\ref{etahatnr}), we obtain the
correct SM values ($\hat{\eta}_\mathrm{NR}\approx 0.6$ \cite{hs13})
if we use $\Gamma\sim 10^{-2}T$.

\subsubsection{Ultra-relativistic limit}

For a wall which has reached ultra-relativistic velocities with very
large gamma factor, several approximations are justified \cite{bm09}.
In the frame of the wall, plasma particles always have enough  energy
to surpass the wall, and the reflection coefficients are
exponentially suppressed. Incoming particles have received no signal
that the wall is approaching and are in equilibrium. Besides,
interactions between plasma particles are time delayed and can be
neglected. Thus, the occupancies evolve undisturbed. As a
consequence, only the equilibrium occupancies of the symmetric phase
are needed in the calculation.   To lowest order in $1/\gamma_w$, the
net force per unit area acting on the wall is given by \cite{bm09}
\begin{equation}
\frac{F}{A}=V(\phi_+)-V(\phi_-)-\sum_i g_i[m^2_i(\phi_-)-m^2_i(\phi_+)]
\int\frac{d^3p}{(2\pi)^32E_{i+}}f^{\mathrm{eq}}_{i+}(p). \label{totalforce0}
\end{equation}
This force does not depend on the wall velocity. Therefore, if it is
positive the wall will accelerate indefinitely. On the other hand, if
$F/A$ is negative, then the wall in fact cannot reach the
ultra-relativistic regime.

Notice that the result (\ref{totalforce0}) can be obtained directly
from Eq. (\ref{eqmicro}), by evaluating the momentum integral in the
$+$ phase. Thus, it does not require the decomposition into
equilibrium occupancies and deviations. We are interested in such a
decomposition, though. More precisely, we intend to decompose the
total force into a driving force and a friction force. This will help
us to introduce, in the next section, a phenomenological model for
the friction which interpolates between the non-relativistic and the
ultra-relativistic limits. Adding and subtracting in Eq.
(\ref{totalforce0}) the finite-temperature correction
$V_{T_+}(\phi_+)$ for the $+$ phase, we have
\begin{equation}
\frac{F}{A}=
{\mathcal{F}}(\phi_+,T_+)-\tilde{\mathcal{F}}(\phi_-,T_+),
\label{totalforce}
\end{equation}
where $\tilde{\mathcal{F}}(\phi_-,T_+)$ is the mean field effective
potential, obtained by keeping only the quadratic terms in a Taylor
expansion of $V_T$ about the $+$ phase \cite{bm09,ekns10},
\begin{equation}
\tilde{\mathcal{F}}(\phi_-,T_+)=V(\phi_-)+V_{T_+}(\phi_+)+\sum_i
[m^2_i(\phi_- )-m^2_i(\phi_+)] \left. \frac{dV_T}{dm^2_i}\right|_{+}.
\label{ftilde}
\end{equation}
In contrast to the non-relativistic case, the total force in
(\ref{totalforce}) does not have a velocity-dependent term, which
could be identified as a friction force. Nevertheless, we can still
decompose the force into a part which comes from the equilibrium
distributions (the driving force) and a part coming from the
deviations (the friction force). The particles in front of the wall
are in equilibrium at temperature $T_+$ and pressure
$p_+(T_+)=-\mathcal{F}(\phi_+,T_+)$. The departures from equilibrium
occur at the wall and behind. The corresponding friction force must
come from the term $-\tilde{\mathcal{F}}(\phi_-,T_+)$ in Eq.
(\ref{totalforce}). We only need to isolate the equilibrium part.

The problem with such a decomposition is that it is not clear which
would be the temperature behind the wall. In the stationary case the
temperature $T_-$ can be calculated from Eqs.
(\ref{disc1}-\ref{disc2}) and is different from $T_+$. In the limit
$v_w\to 1$ (i.e., for a very fast detonation) we still obtain
$T_-\neq T_+$. Nevertheless, there is no reason to assume that the
stationary solution should match the runaway solution. Notice that
the only temperature appearing in Eqs.
(\ref{totalforce}-\ref{ftilde}) is $T_+$, suggesting that, in the
runaway case, we have $T_-=T_+$. Indeed, for stationary solutions the
released latent heat goes into bulk motions of the fluid and
reheating of the plasma, whereas for runaway solutions the energy
goes mainly into accelerating the wall. Assuming $T_-=T_+$, the
occupancies in each phase (in the plasma frame) are given by
\begin{eqnarray}
f_{i+}(p_+)&=&1/[\exp(E_{i+}/T_+)\mp 1],\ \\
f_{i-}(p_-)&=&1/[\exp(E_{i-}/T_+)\mp
1]+\delta f_i, \label{decompfrun}
\end{eqnarray}
with $E^2_{i\pm}=p_{\pm}^2+m_{i\pm}^2$.

In fact, we can calculate the exact form of the occupancies just
behind the wall. After the passage of the wall the occupancies
$f_{i+}(p_+)$ are undisturbed, but the energy and momentum of a
particle have changed. In the plasma frame, these changes are given
by \cite{bm09}
\begin{equation}
E_{i-}-E_{i+}=p_--p_+=\frac{m_{i-}^2-m_{i+}^2}{2(E_{i-}-p_-)}.
\end{equation}
We thus have ${E_{i+}(p_+)}={E_{i-}(p_-)}+
(m_{i+}^2-m_{i-}^2)/[2(E_{i-}-p_{-})]$, and
\begin{equation}
f_{i-}(p_-)=\left[\exp\left(\frac{E_{i-}(p_-)}{T_+}+
\frac{m_{i+}^2-m_{i-}^2}{2(E_{i-}-p_{-})T_+}\right)\mp 1\right]^{-1}.
\label{firun}
\end{equation}
It seems natural to decompose the occupancies (\ref{firun}) in the
form (\ref{decompfrun}),  which can be taken as the definition of
$\delta f_i$. If, e.g., the mass difference $m_{i+}-m_{i-}$ is small,
then the deviation $\delta f_i$ will be small too.

According to Eqs. (\ref{decompfrun}) and (\ref{firun}), the
deviations from equilibrium vanish only in front of the bubble wall,
in contrast with the non-relativistic case, in which the deviations
vanish also behind the wall. Indeed, the $(\phi')^2$ factor in Eq.
(\ref{eta}) indicates that the deviations are localized at the bubble
wall. The difference arises because, in the UR case, the wall has
passed so quickly that there was no time for the plasma to recover
the equilibrium. Notice, anyway, that Eq. (\ref{firun}) gives the
occupancies behind the wall but close to it. At some distance behind
the wall the plasma will reach the equilibrium (and some reheating
will occur).

It is easy to obtain the macroscopic version of the decomposition
(\ref{decompfrun}), directly from Eq. (\ref{totalforce}). For
$T_-=T_+$, the driving force is given by
\begin{equation}
\frac{F_{\mathrm{dr}}}{A}=
{\mathcal{F}}(\phi_+,T_+)-{\mathcal{F}}(\phi_-,T_+).
\end{equation}
Therefore, the departure from equilibrium causes a friction force
given by
\begin{equation}
\frac{F_{\mathrm{fr}}}{A}={\mathcal{F}}(\phi_-,T_+)-
\tilde{\mathcal{F}}(\phi_-,T_+),
\label{fricpmen}
\end{equation}
Since $v_w \simeq 1$, we can define an ultra-relativistic friction
coefficient $\eta_\mathrm{UR}$ by
\begin{equation}
\eta_\mathrm{UR}=-{F_{\mathrm{fr}}}/{A}, \label{fricurmicro}
\end{equation}
so that the total force is given by
\begin{equation}
\frac{F}{A}=\left.\left(p_--p_+\right)\right|_{T_+}-\eta_\mathrm{UR}v_w.
\label{decompf}
\end{equation}

Notice that we may also write Eq. (\ref{totalforce}) as
$F/A=\tilde{\mathcal{F}}(\phi_+,T_+)-\tilde{\mathcal{F}}(\phi_-,T_+)$,
since $\tilde{\mathcal{F}}(\phi_+,T_+)=\mathcal{F}(\phi_+,T_+)$. If
this mean-field potential difference is negative, then the bubble
wall cannot run away. For instance, the wall never runs away in a
``fluctuation induced'' first-order phase transition \cite{bm09},
i.e., a phase transition which is first-order due to the thermal part
of the effective potential, $V_T$. For example, consider the well
known high-temperature expansion of Eq. (\ref{f1loop}),
\begin{equation}
V_T(\phi)=-\sum_{i}\frac{\pi^2g_ic'_iT^4}{90}+
\sum_{i}g_ic_i\frac{T^2m^2_i(\phi)}{24}
-\sum_{\mathrm{bosons}}g_i\frac{Tm_i^3(\phi)}{12\pi}
+\mathcal{O}(m^4), \label{vpot}
\end{equation}
which is valid for many models. Here, $c_i=1$ ($1/2$) and $c'_i=1$
($7/8$) for bosons (fermions). Only bosons contribute to the $m_i^3$
terms.  In fact, after the resummation of daisy diagrams, only the
transverse polarizations of gauge bosons remain in general in this
contribution. The cubic term in Eq. (\ref{vpot}) is very important
since it may cause a barrier in the free energy and, thus, a
first-order phase transition. This term is not present in the mean
field potential. If the first-order character of the phase transition
is due to this term alone, then in the mean field potential the
minimum $\phi_-$ will raise above the minimum $\phi_+$. As a
consequence, the net force (\ref{totalforce}) will be negative and
the wall will reach a terminal velocity.

Let us  consider the friction coefficient $\eta_\mathrm{UR}$ for the
case $m/T\ll 1$ (as we did for $\eta_\mathrm{NR}$). According to Eq.
(\ref{vpot}) we have, to lowest order in $m/T$,
\begin{equation}
\eta_\mathrm{UR}\approx \sum_{\mathrm{bosons}}  \frac{g_iT}{12\pi}
\left[m_i^3(\phi_-)-m_i^3(\phi_+)\right].
\label{fricurm}
\end{equation}
We have assumed that the contribution of gauge bosons is important.
Otherwise, the $\mathcal{O}(m^4)$ terms should be considered. The
generalization is straightforward. Let us further simplify the
problem by specifying to the case $\phi_+=0$ and assuming particle
masses of the form $m_i(\phi)=h_i\phi$. Then, we have
\begin{equation}
\eta_\mathrm{UR}\approx \sum_{\mathrm{bosons}} \frac{g_i h_i^3 }{12\pi}T\phi_0^3
\equiv \hat{\eta}_{\mathrm{UR}}T\phi_0^3, \label{fricur}
\end{equation}
where we used again the notation $\phi_0$ for the symmetry-braking
minimum, and we have defined a dimensionless coefficient
$\hat{\eta}_{\mathrm{UR}}$ which contains the parameters of the
model.

\subsection{Runaway conditions}

According to Eqs. (\ref{fuerzas}) and (\ref{decompf}), in the runaway
regime we have
\begin{equation}
\sigma \ddot{z}_w=p_-(T_+)-p_+(T_+)-\eta_{\mathrm{UR}}, \label{prenecessary}
\end{equation}
which gives the necessary condition for the wall to run away,
\begin{equation}
p_-(T_+)-p_+(T_+)>\eta_{\mathrm{UR}}. \label{necessary}
\end{equation}
This condition does not depend on the non-relativistic friction
parameter $\eta_{\mathrm{NR}}$ and is equivalent to
$\tilde{\mathcal{F}}(\phi_+,T_+)-\tilde{\mathcal{F}}(\phi_-,T_+)>0$,
which is the criterion provided in Ref. \cite{bm09}. We wish to
emphasize that Eq. (\ref{necessary}) is just a \emph{necessary}
condition: if it is not fulfilled, then the wall cannot run away
(since the total force in the runaway regime would then be negative).
However, it is not a \emph{sufficient} condition for the wall to run
away, since it was obtained by assuming that the wall is already in
the runaway regime. Thus, Eq. (\ref{necessary}) implies that the
runaway solution exists, but does not guarantee that it will be
realized. It is well known that multiple hydrodynamical solutions may
exist for the same set of parameters. Before the wall reaches
ultra-relativistic velocities, the forces are not like in Eq.
(\ref{prenecessary}), and it may be possible for the friction to
compensate the driving force. Even if the friction force is smaller
for smaller velocities, hydrodynamic effects slow down the wall,
acting effectively as a friction \cite{ms09}.

To be specific, consider a very fast detonation. In the limit in
which the terminal velocity $v_w\to 1$, we have $v_{\pm}\to 1$.
However, for a detonation, the difference of temperature between both
sides of the wall does not disappear in the limit $v_w\to 1$. The
friction force is not too sensitive to temperature gradients, though,
and it should be given, for $\gamma_w\gg1$, by
$F_{\mathrm{fr}}/A=\eta_{\mathrm{UR}}$. Thus, the friction for the
detonation is as in Eq. (\ref{prenecessary}). However, the driving
force is not, since it is very sensitive to temperature gradients.
Using the approximation (\ref{draprox}), we have
$F_{\mathrm{dr}}/A=p_-(T_-)-p_+(T_+)-\langle s\rangle (T_--T_+)$. For
a stationary solution, the driving force equals the friction force,
and we have $F_{\mathrm{dr}}/A=\eta_{\mathrm{UR}}$. For smaller
velocities, the friction is in principle smaller, and so is the
driving force. Hence, for detonations,
$F_{\mathrm{dr}}/A=\eta_{\mathrm{UR}}$ is the maximum value the
driving force can reach. We thus have a necessary condition for the
stationary solution, namely, $F_{\mathrm{dr}}/A
\leq\eta_{\mathrm{UR}}$. If this condition is not satisfied, i.e., if
\begin{equation}
p_-(T_-)-p_+(T_+)-\langle s\rangle (T_--T_+)>\eta_{\mathrm{UR}},
\label{sufficient}
\end{equation}
then the stationary detonation solution does not exist.

We notice that the two conditions (\ref{necessary}) and
(\ref{sufficient}) are really different\footnote{According to the
exact expression (\ref{dr}), for $T_-\neq T_+$  we could still have
$F_{\mathrm{dr}}/A=p_-(T_+)-p_+(T_+)$ if $\phi(z)\equiv\phi_-$ across
the wall, which is not the case.} since, for the detonation, the
temperature $T_-$ is always higher than $T_+$. If Eq.
(\ref{sufficient}) is fulfilled, then the necessary condition
(\ref{necessary}) is also fulfilled. Indeed, given
$\eta_{\mathrm{UR}}$ and $T_+$, the force $p_-(T_+)-p_+(T_+)$ is
larger than $p_-(T_-)-p_+(T_+)-\langle s\rangle (T_--T_+)$, since the
latter is affected by hydrodynamics. Therefore, Eq.
(\ref{sufficient}) gives a \emph{sufficient condition} for runaway
walls, corresponding to the detonation velocity becoming $v_w=1$.

According to the above, as the parameters are varied there will exist
a runaway solution before the stationary solution ceases to exist,
and vice versa. Hence, the runaway and the detonation solutions will
coexist in a certain range of parameters. This range is delimited by
the conditions (\ref{necessary}) and (\ref{sufficient}). Stability
analysis indicate that the detonation is generally a stable solution
\cite{unstable}. Therefore, we expect that the wall will not runaway
in the coexistence range. This range will be short if the latent heat
is relatively small ($L\ll T^4$). Indeed, in such a case we will have
$T_+\simeq T_-$, and Eqs. (\ref{necessary}) and (\ref{sufficient})
will give a similar condition.

\section{Phenomenological model for the friction} \label{runaway}

In order to simplify the treatment of phase transition fronts, it is
usual to replace the last term in the field equation
(\ref{eqmicro2}), corresponding to the deviations from equilibrium,
with a phenomenological damping term.

\subsection{Existing models}

\subsubsection{``Old'' phenomenological model}

A widely used approximation is a damping term proportional to
$u^{\mu}\partial _{\mu }\phi $. The coefficient of this term can be
determined by comparing the resulting friction force (or wall
velocity) with the results of microphysics calculations such as those
considered in the previous section. In the general case, the
coefficient is field-dependent,
\begin{equation}
\partial _{\mu }\partial ^{\mu }\phi +\frac{\partial \mathcal{F}}{\partial
\phi }+f(\phi)\,  u^{\mu }\partial _{\mu }\phi=0.
\label{eqfieldnr}
\end{equation}
We shall refer to this approach as the ``old'' phenomenological
model.

Proceeding like in the previous section, i.e., assuming a field
profile of the form $\phi(z,t)=\phi[\gamma_w(z-z_w)]$, considering a
reference frame which instantaneously moves with the wall,
multiplying by $\phi ^{\prime }$ and integrating across the wall, we
obtain
\begin{equation}
\sigma \ddot{z}_w=\int \frac{\partial{\mathcal{F}}}{\partial \phi}\frac{d\phi}{dz}dz+
\int \gamma v f(\phi)(\phi')^2dz. \label{fenonr}
\end{equation}
where $v(z)$ is the (negative) fluid velocity and
$\gamma=1/\sqrt{1-v^2}$. The first term in the rhs is the driving
force, and the last term gives a friction force which, for small
velocities, can reproduce the form (\ref{eta}). Indeed, for small
$v_w$ the temperature and fluid velocity are not significantly
altered by the wall. Hence, the temperature is homogeneous and the
fluid velocity is given by $v\approx -v_w$. We thus have
\begin{equation}
\sigma \ddot{z}_w=p_{-}(T)-p_{+}(T)- \eta_{\mathrm{NR}}v_w. \label{fenonrlimit}
\end{equation}
with
\begin{equation}
\eta_{\mathrm{NR}}=\int f(\phi)\phi'^2dz \label{etanr}
\end{equation}
The function $f(\phi)$ can be chosen so that Eq. (\ref{etanr})
reproduces the friction coefficient obtained from a microphysics
calculation, such as Eq. (\ref{eta}). For the case of  small $m_i/T$,
we can choose
\begin{equation}
f(\phi)=\hat{\eta}_{\mathrm{NR}}\phi^2/T, \label{funcf}
\end{equation}
so that we obtain the result of Eq. (\ref{etatth}).

In the general case, hydrodynamics must be taken into account. In
particular, Eqs. (\ref{disc1}-\ref{disc2}) imply that the fluid
velocity and the temperature are in general different on each side of
the wall. In sec. \ref{drf} we found an approximation for the first
integral in Eq. (\ref{fenonr}). For the second integral, we notice
that $\phi'(z)^2$ peaks inside the wall and vanishes outside.
According to the weighted mean-value theorem for integrals
\cite{apostol}, we can replace the function $f(\phi)\gamma v$ with
its value at a certain point $z=\bar{z}$ inside the wall. We shall
approximate this value by $f(\phi_0/2)\langle\gamma v\rangle$, with
$\langle\gamma v\rangle\equiv (\gamma_-v_-+\gamma_+v_+)/2$. With this
approximation, the friction force per unit area, including
hydrodynamics effects, is given by
\begin{equation}
\frac{F_{\mathrm{fr}}}{A}=\eta_{\mathrm{NR}}\langle\gamma v\rangle .
\label{fricnrhidro}
\end{equation}
For the case of Eq. (\ref{funcf}), we have
$\eta_{\mathrm{NR}}=\hat{\eta}_{\mathrm{NR}} (\phi_0^2/4)\,\sigma/T$.

Although the phenomenological model (\ref{eqfieldnr}) gives the
correct form of the friction in the non-relativistic limit, for
relativistic velocities the friction force (\ref{fricnrhidro}) is of
the form $F_{\mathrm{fr}}/A\sim - v_w\gamma_w$, which does not
saturate in the ultra-relativistic limit. As a consequence, this
model gives always stationary solutions. Even for
${\eta}_{\mathrm{NR}}\to 0$ we have detonation solutions and never
runaway solutions.

\subsubsection{A friction which saturates}

Recently \cite{ekns10}, a simple modification of the usual damping
term $u^{\mu}\partial_{\mu}\phi$, was considered in order to take
into account the friction saturation. In our notation, the damping
term is of the form
\begin{equation}
\frac{f(\phi)\, u^{\mu}\partial_{\mu}\phi}{\sqrt{1+
(\lambda_{\mu}u^{\mu})^2}}. \label{replekns}
\end{equation}
The constant vector $\lambda_{\mu}$ is given by
$\lambda_{\mu}=(0,0,0,1)$ in the wall frame, and the function
$f(\phi)$ is given by Eq. (\ref{funcf}),
$f(\phi)=(\hat{\eta}_{\mathrm{NR}}/T)\phi^2$, so that the friction
parameter can be determined by comparison with microphysics
calculations in the non-relativistic regime. With the modification
(\ref{replekns}), the factor $\gamma v$ in Eq. (\ref{fenonr}) gets
replaced by $v$. This gives a friction per unit area of the form
\begin{equation}
F_{\mathrm{fr}}/A=\eta_{\mathrm{NR}} \langle v\rangle, \label{fricurhidro}
\end{equation}
which has a correct small-velocity behavior as well as Eq.
(\ref{fricnrhidro}) and, besides, saturates for large $\gamma v$.
More recently, in Ref. \cite{hs13} the old model  was considered,
again with a function $f(\phi)$ of the form $f(\phi)=(\eta/T)\phi^2$,
but with a velocity-dependent coefficient $\eta= \eta_0/\gamma$, so
that the $\gamma$ factor in Eq. (\ref{fenonr}) cancels out. The
coefficient $\eta_0$ is determined by calibrating with
non-relativistic results. Therefore, this approach gives again a
friction of the form (\ref{fricurhidro}).

Although the friction (\ref{fricurhidro}) saturates for $v\to 1$, we
note that this approximation is  too simplistic, as the friction
coefficient is determined in the NR limit. Numerically, this model
cannot give a correct value for the friction in the UR limit, unless
$\eta_{\mathrm{UR}}={\eta}_{\mathrm{NR}}$. Conversely, if the
coefficient in Eq. (\ref{fricurhidro}) were determined by
$\eta_{\mathrm{UR}}$ instead of $\eta_{\mathrm{NR}}$, then the model
would not give the correct value of the friction in the
non-relativistic limit. Notice that $\eta_{\mathrm{NR}}$ and
${\eta}_{\mathrm{UR}}$ will have in general different parametric
dependence, given, e.g., by Eqs. (\ref{etatth}-\ref{etahatnr}) or by
Eq. (\ref{fricur}), respectively. It is clear  that a model with a
single parameter falls short of describing the friction in the two
opposite regimes. A more realistic interpolating model should include
two free parameters.

\subsection{Adding a free parameter}

The  vector  $\lambda_{\mu}$ in the model (\ref{replekns}) may arise
as an effective value of a covariant vector. It is associated to the
presence of the wall and, therefore, should depend on gradients of
the field.  Therefore, we propose the following phenomenological
equation for $\phi$,
\begin{equation}
\partial _{\mu }\partial ^{\mu }\phi +\frac{\partial \mathcal{F}}{\partial
\phi }+  \frac{f(\phi)\,\partial_{\mu}\phi\,u^{\mu}}{\sqrt{1+
\left[g(\phi)\,\partial_{\mu}\phi \,u^{\mu}\right]^2}}=0.
\label{eqfield}
\end{equation}%
We have chosen a form of the damping term similar to that of Eq.
(\ref{replekns}), which will allow us to compare our results to those
of Refs. \cite{ekns10,hs13}. Notice that other forms [e.g.,
${u^{\mu}\partial_{\mu}\phi}/({{1+ \lambda_{\mu}u^{\mu}}})$]
reproduce as well the two behaviors $F_{\mathrm{fr}}\sim v$ and
$F_{\mathrm{fr}} \sim\mathrm{constant}$ in the NR and UR limits,
respectively. Thus, in our model the phenomenological vector
$\lambda_{\mu}$  arises from derivatives of the field. In practice,
the important difference with the model (\ref{replekns}) is that the
$z$ component $\lambda_{z}$ is not set to $1$, but gives a new free
parameter  which can be set so as to obtain the correct UR limit.

Let us consider a reference frame which instantaneously moves with
the wall (in which most time-derivatives vanish). The
phenomenological damping term is given by
\begin{equation}
\frac{f(\phi)\,\phi'\,\gamma v}{\sqrt{1+
\left[g(\phi)\,\phi' \,\gamma v\right]^2}}=0.
\label{eqfieldz}
\end{equation}%
The field derivative appearing in the denominator is responsible for
a correct correspondence  with the  deviations from local
equilibrium, as discussed in section \ref{micro}. In the
non-relativistic limit, the damping term (\ref{eqfieldz}) is
proportional to $\phi'$ and, thus, is localized inside the wall. This
corresponds to assuming that the departure from equilibrium occurs
inside the wall. This is reasonable for a slow wall since, after the
wall has passed through a given point in space, the system quickly
reaches the thermodynamical equilibrium. On the other hand, in the
ultra-relativistic limit  $\phi'$ cancels out (together with $\gamma
v$). For $f(\phi)\propto \phi^2$ as in Eq. (\ref{funcf}), the damping
term vanishes in front of the wall but not behind it, since $\phi$
varies from $\phi_+=0$ to $\phi_-\neq 0$. This is also reasonable
since, as we have seen, in the runaway regime the deviations from
equilibrium occur at the wall and behind.

Proceeding as before, in the reference frame which moves with the
wall we obtain
\begin{equation}
\sigma \ddot{z}_w=\int \frac{\partial{\mathcal{F}}}{\partial \phi}
\frac{d\phi}{dz}dz+
\int \frac{\gamma v f(\phi)(\phi')^2}{ \sqrt{1+
(\gamma v)^2g(\phi)^2 (\phi')^2}}dz. \label{feno}
\end{equation}
The friction is given by the second integral in Eq. (\ref{feno}). For
$v_w\ll 1$ we recover Eqs. (\ref{fenonr}-\ref{fenonrlimit}), i.e., we
obtain a linear friction force,
\begin{equation}
\left.\frac{F_{\mathrm{fr}}}{A}\right|_{\mathrm{NR}}=-v_w
 \int dz f(\phi)(\phi')^2\equiv -{\eta}_{\mathrm{NR}}v_w, \label{ffrnr}
\end{equation}
whereas for $\gamma v\gg 1$ we obtain a constant friction force,
\begin{equation}
\left.\frac{F_{\mathrm{fr}}}{A}\right|_{\mathrm{UR}}=
-\int \frac{f(\phi)}{
 g(\phi) }d\phi\equiv -{\eta}_{\mathrm{UR}}. \label{ffrur}
\end{equation}
The functions $f(\phi)$ and $g(\phi)$ can be chosen so that Eqs.
(\ref{ffrnr}) and (\ref{ffrur}) give the desired values of the
friction coefficients ${\eta}_{\mathrm{NR}}$ and
${\eta}_{\mathrm{UR}}$ in different models. For instance, a function
$f(\phi)=\hat{\eta}_{\mathrm{NR}}{\phi^2}/{T}$ gives again a friction
coefficient of the form (\ref{etatth}) in the non-relativistic limit.
With this choice for $f(\phi)$, a $\phi$-independent function
\begin{equation}
g(\phi)=\frac{\hat{\eta}_{\mathrm{NR}}}{\hat{\eta}_{\mathrm{UR}}}\frac{1}{3T^{2}}
\label{funcg}
\end{equation}
gives a friction coefficient of the form (\ref{fricur}) in the
ultra-relativistic limit. For intermediate velocities the friction
depends on the velocity profile as well as on the field profile.

Solving the set of differential equations for $v$, $T$ and $\phi$ is
out of the scope of the present paper. Instead of that, we shall use
the thin-wall approximation as before in order to obtain analytical
results. Invoking again the weighted mean-value theorem for
integrals, we notice that the friction integral in  Eq. (\ref{feno})
can be evaluated by replacing the whole coefficient of $(\phi')^2$
with its value at a certain  point $z=\bar{z}$ inside the
wall\footnote{Although the damping term  (\ref{eqfieldz}) does not
vanish behind the wall in the UR limit, in Eq. (\ref{feno}) there is
an extra factor of $\phi'$ in the integral for the force acting on
the wall.}. It should be a good approximation to assume that
$\phi'(z)$ is symmetric and, thus, picks the values $\phi=\phi_0/2$
and $\lambda_z=g(\phi_0/2)\phi'(\bar{z})$. In the thin-wall
approximation one can estimate $\phi'$ as, e.g.,
$\phi'\sim\phi_0/l_w$. In fact, whatever approximation we use for
$\phi'$, we still have the freedom to choose the function $g(\phi)$
so as to obtain the desired value of the friction parameter
$\eta_{\mathrm{UR}}$. With these approximations the model essentially
reduces to Eq. (\ref{replekns}), only that in our case $\lambda_z\neq
1$. It remains to find an approximation for $v(\bar{z})$. The
simplest and more reasonable one would be $v(\bar{z})\approx \langle
v\rangle \equiv (v_++v_-)/2$. This gives a friction force
\begin{equation}
\frac{F_{\mathrm{fr}}}{A}=\frac{\eta_{\mathrm{NR}}\langle v\rangle}{\sqrt{1-
(1-\lambda_z^2)\langle v\rangle ^2}} .  \label{fricmejor}
\end{equation}%
Notice that in  Eq. (\ref{fricnrhidro}) the approximation $\langle
v\gamma\rangle$ was used instead of $\langle v\rangle \gamma(\langle
v\rangle)$. In order to compare with previous results, in the present
paper we shall use, accordingly,
\begin{equation}
\frac{F_{\mathrm{fr}}}{A}=\eta_{\mathrm{NR}}\left\langle\frac{v }{ \sqrt{1-
(1-\lambda_z^2)v^2}}\right\rangle  \label{fric}
\end{equation}%
instead of Eq. (\ref{fricmejor}), so that for $\lambda_z=0$ we
recover Eq. (\ref{fricnrhidro}). For $\lambda_z=1$ we recover Eq.
(\ref{fricurhidro}). There should not be a  significant difference in
using either of Eqs. (\ref{fricmejor}), (\ref{fric}).

\begin{figure}[bt]
\centering
\epsfysize=7cm \leavevmode \epsfbox{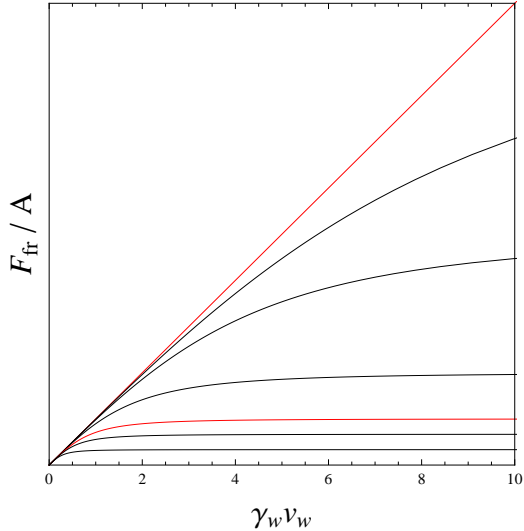}
\caption{A friction force of the form $v_w/\sqrt{1-(1-\lambda_z^2)v_w^2}$.
From top to bottom the curves correspond to $\lambda=0,0.1,0.2,0.5,1,1.5$ and $3$.
Red lines indicate the cases $\lambda=0$ and $\lambda=1$.}
\label{figforce}
\end{figure}

The behavior of this friction force for different values of
$\lambda_z$ is shown in Fig. \ref{figforce} (where the effects of
hydrodynamics have been neglected, so that $v=-v_w$).
The old model (\ref{eqfieldnr}) corresponds to the case $\lambda_z=0$
and gives a friction  which never saturates. We see that, the larger
the value of $\lambda_z$, the sooner the friction saturates. The
models considered in Refs. \cite{ekns10,hs13} correspond to  the case
$\lambda_z=1$. This gives a friction  which, in the
ultra-relativistic limit, is a constant proportional to
$\eta_{\mathrm{NR}}$.  In the limit $v_w\to 1$, our model gives
$F_{\mathrm{fr}}/A=- \eta_{\mathrm{NR}}/\lambda_z$. Comparing with
Eq. (\ref{fricurmicro}), we see that the parameter $\lambda_z$ is
given by
\begin{equation}
\lambda_{z}=\frac{\eta_\mathrm{NR}}{\eta_\mathrm{UR}}, \label{lambda}
\end{equation}
which shows explicitly that the particular case $\lambda_z=1$
corresponds to ${\eta_\mathrm{UR}}={\eta_\mathrm{NR}}$, whereas the
case $\lambda_z=0$ corresponds to the limit
${\eta_\mathrm{UR}}\to\infty$. In terms of the two independent
friction parameters, the equation for the wall reads
\begin{equation}
\sigma \ddot{z}_w=p_{-}(T_-)-p_{+}(T_+) +
\langle s\rangle\left( T_{+}-T_{-}\right)
+\left\langle\frac{\eta_{\mathrm{NR}}\eta_{\mathrm{UR}}v }{
\sqrt{\eta_{\mathrm{NR}}^2v^2+\eta_{\mathrm{UR}}^2(1-v^2)}}\right\rangle ,
\label{eqwall}
\end{equation}%
where we have used the approximation (\ref{draprox}) for the driving
force.

\section{The  wall velocity} \label{result}

In this section we shall study the wall velocity, considering
independent variations of the friction coefficients, the nucleation
temperature $T_n$, and other thermodynamical quantities. To proceed
further, we need to consider an equation of state. It is convenient
to use the well-known bag EOS, which is simple enough to obtain
analytical and model independent results.

\subsection{The bag EOS} \label{bag}

The bag EOS can be derived from a free energy density of the form
\begin{equation}
\mathcal{F} _{+}\left( T\right) =-a_{+}T^{4}/3+\epsilon, \ \
\mathcal{F}_{-}\left( T\right) =-a_{-}T^{4}/3.  \label{eos}
\end{equation}
Thus, the metastable phase is characterized by radiation and false
vacuum, whereas in the stable phase we only have radiation. In this
model the speed of sound is a constant, $c_{\pm}=1/\sqrt{3}$. More
generally, we could consider a nonvanishing vacuum energy density
$\epsilon_-$ in the stable phase. For simplicity we just set
$\epsilon_-=0$ and $\epsilon_+=\epsilon$. The only difference is that
all our results below would otherwise depend\footnote{The false
vacuum energy density  affects the development of a phase transition
through the dependence of the expansion rate of the universe on the
total energy density. However, our results for the wall velocity as a
function of thermodynamic and friction parameters depend only on
$\Delta \epsilon$.} on $\Delta\epsilon=\epsilon_+-\epsilon_-$ instead
of $\epsilon$. Some words are worth, though, on using these results.
In a real model, $\epsilon_+$ and $\epsilon_-$ will be given in
general by the scale $v$ of the model. Thus, we will naturally have
$\epsilon_{\pm}\sim v^4$. In contrast, $\Delta\epsilon$ may be much
smaller than that, since only a part of the false vacuum energy is in
general released at $T=T_c$. For instance, in a second-order phase
transition we will have $\Delta\epsilon =0$. Hence, for application
of our results, it must be taken into account that $\Delta
\epsilon\equiv \epsilon$ corresponds to the false vacuum energy
density \emph{that is released at $T=T_c$}, and not the \emph{total}
false vacuum energy density. For that reason, it is convenient to
use, instead of $\epsilon$, the latent heat $L$, which in a
particular model can be easily calculated as explained in section
\ref{hydro}. Notice that $L$ is larger than $\epsilon$, since thermal
energy is released in addition to false vacuum energy. For the bag
EOS the critical temperature is given by the equation
\begin{equation}
(a_+-a_-)T_c^4=3\epsilon , \label{tcbag}
\end{equation}
and we have the simple relation
\begin{equation}
L=4\epsilon , \label{lbag}
\end{equation}
since a thermal energy density $(a_+-a_-)T_c^4$ is released in
addition to the vacuum energy density $\epsilon$.

We shall consider bubbles nucleated at a temperature $T_n<T_c$.
Hydrodynamics will be dominated by  the pressure difference between
phases, $\Delta p\equiv p_-(T_n)-p_+(T_n)$, and by the energy density
which is released at the phase transition fronts,
$\Delta\rho\equiv\rho_+(T_n)-\rho_-(T_n)$.  For $T_n$ close to $ T_c$
we have $\Delta p\simeq 0$ and $\Delta\rho\simeq L$. In contrast, for
$T_n\to 0$ (i.e., for strong supercooling), the pressure difference,
as well as the released energy, are just given by $\Delta\rho=\Delta
p=\epsilon$.  As we have seen, in the general case the driving force
is not just given by $\Delta p$, since it is affected by temperature
gradients.

Let us first consider  stationary solutions. The fluid discontinuity
equations (\ref{disc1}-\ref{disc2}) can be used to eliminate the
pressure difference $p_--p_+$ from Eq. (\ref{eqwall}). For the bag
EOS, this gives a relatively simple equation \cite{ms12,ms09},
\begin{equation}
\frac{2v_{+}v_{-}}{1-3v_{+}v_{-}}-\frac{(
1+\hat{s}
) ( 1-
\hat{T}
)}{3\alpha_+} +
\frac{\eta_{\mathrm{NR}}
}{L}\left[ \frac{|v_+| \gamma_+}{ \sqrt{1+ \lambda_z^2
(v_+\gamma_+)^2}}+\left( +\to -\right)\right] =0,  \label{fricbag}
\end{equation}
where
\begin{equation}
\alpha_+ =\frac{\epsilon}{
a_{+}T_+^{4}}=\frac{L}{4 a_{+}T_+^{4}},
\end{equation}
\begin{equation}
\hat{s}\equiv\frac{s_{-}}{s_{+}}=\frac{a_-}{a_+}\hat{T},
^{3}
\end{equation}
and
\begin{equation}
\hat{T}\equiv\frac{T_{-}}{T_{+}}=\left[\frac{a_+}{a_-} \left(
1- \frac{1+v_{+}v_{-}}{1-3v_{+}v_{-}}3\alpha_+\right) \right]
^{1/4}.  \label{tmatme}
\end{equation}%
In the last expression, we used again the fluid equations
(\ref{disc1}-\ref{disc2}) to write the enthalpy ratio $w_-/w_+$ in
terms of $v_+$ and $v_-$. Thus,  Eq. (\ref{fricbag}) relates the
fluid variables $v_+,v_-,\alpha_+$. Besides these variables, Eq.
(\ref{fricbag}) only depends on the parameter ratios
$\eta_{\mathrm{NR}}/L,\eta_{\mathrm{UR}}/\eta_{\mathrm{NR}},a_-/a_+$.
The later can be written as $a_-/a_+=1-3\alpha_c$, with
$\alpha_c=\epsilon/(a_{+}T_c^{4})$. Additionally, Eqs.
(\ref{disc1}-\ref{disc2}) give a relation between $v_+$ and $v_-$
\cite{s82},
\begin{equation}
v_{+}=\frac{1}{1+\alpha_+}\left[ {\frac{1}{6v_{-}}+
\frac{v_{-}}{2}\pm \sqrt{\left( \frac{1}{6v_{-}}
+\frac{v_{-}}{2}\right) ^{2}+\alpha_+^{2}+\frac{2}{3}\alpha_+-\frac{1}{
3}}}\right] . \label{vmavme}
\end{equation}%
The plus and minus signs in Eq. (\ref{vmavme}) indicate that we have
two hydrodynamical solutions, namely, detonations and deflagrations.
Using this relation, one can readily solve Eq. (\ref{fricbag}) to
obtain the velocities $v_{\pm}$ as functions of $\alpha_+$ (hence, as
functions of $T_+$). To obtain the wall velocity as a function of the
nucleation temperature $T_n$, appropriate boundary conditions must be
used. The relation between the fluid velocities $v_{\pm}$ and the
wall velocity $v_w$, as well as the relation between $\alpha _{+}$
and  $\alpha _{n}\equiv \epsilon/(a_{+}T_n^{4})$, depend on the type
of hydrodynamic solution. For detonations we have simply $\alpha
_{+}=\alpha _{n}$ and $v_+=-v_w$. For deflagrations, the matching
conditions at the shock discontinuity must be used. These give
\begin{equation}
\frac{v_{w}-|v_{+}|}{1-|v_{+}|v_{w}}=\frac{\sqrt{3}\left( \alpha _{n}-\alpha_+
\right) }{\sqrt{\left( 3\alpha _{n}+\alpha_+ \right) \left( 3\alpha_+
+\alpha _{n}\right) }}.  \label{vfl}
\end{equation}%
For traditional deflagrations we have $v_{-}=-v_{w}$, whereas for
Jouguet deflagrations we have $v_{-}=c_s=-1/\sqrt{3}$ (for details
see Refs. \cite{ms12,ms09}).

Using the bag EOS, we can also express the runaway conditions in
terms of the thermodynamic parameters and the ultra-relativistic
friction coefficient. From Eqs. (\ref{eos}-\ref{lbag}), we have
$p_-(T_n)-p_+(T_n)=(L/4)(1-T_n^4/T_c^4)$, and the {necessary}
condition (\ref{necessary})  gives simply
\begin{equation}
\frac{T_n^4}{T_c^4}+\frac{\eta_\mathrm{UR}}{\epsilon}<1. \label{necessbag}
\end{equation}
We may express this condition as
$\eta_{\mathrm{UR}}<\eta_{\mathrm{nec}}$, with
\begin{equation}
\eta_{\mathrm{nec}}/\epsilon\equiv 1-\alpha_c/\alpha_n. \label{etanec}
\end{equation}
In terms of the $\alpha$ variables, we have
\begin{equation}
\alpha_n>\alpha_{\mathrm{nec}}\equiv\frac{\alpha_c}{1-\eta_\mathrm{UR}/
\epsilon}.\label{alfanec}
\end{equation}
We remark that Eqs. (\ref{necessbag}-\ref{alfanec}) only give a
necessary condition for the wall to run away, as the runaway solution
may coexist with a detonation solution, which is presumably stable.
Nevertheless, detonations are not possible if the sufficient
condition (\ref{sufficient}) is fulfilled. For the bag EOS and for a
detonation with $v_w\approx 1$,  we have $p_-(T_-)-p_+(T_+)= L/2$,
$T_-/T_+=(a_-/a_+)^{-1/4}(1+3\alpha_n)^{1/4}$, and
$s_-/s_+=(a_-/a_+)^{1/4}(1+3\alpha_n)^{3/4}$ (for analytic
approximations for ultra-relativistic detonations see Ref.
\cite{ms09}). We thus obtain the sufficient runaway condition
$\eta_\mathrm{UR}<\eta_\mathrm{suf}$, with
\begin{equation}
\frac{\eta_\mathrm{suf}}{\epsilon}=\frac{2}{3\alpha_n}
\left[(1-3\alpha_c)^{\frac{1}{4}}
(1+3\alpha_n)^{\frac{3}{4}}- (1-3\alpha_c)^{-\frac{1}{4}}
(1+3\alpha_n)^{\frac{1}{4}}\right]. \label{sufibag}
\end{equation}
Equivalently, for given values of ${\eta_\mathrm{UR}}/{\epsilon}$ and
$\alpha_c$ we may express this condition as
$\alpha_n>\alpha_{\mathrm{suf}}$, i.e., in terms of the amount of
supercooling which is sufficient for the wall to runaway. The value
of $\alpha_{\mathrm{suf}}$ is obtained by inverting Eq.
(\ref{sufibag}). This amounts to solving a quartic equation for the
variable $x=(1+3\alpha_n)^{1/4}$. The expression for
$\alpha_{\mathrm{suf}}$ is cumbersome and we shall not write it down.

Before going on to the interpretation of  these results, it is worth
comparing them with the previous works \cite{ekns10,hs13}. The
runaway necessary condition can be written in terms of fundamental
parameters rather than phenomenological ones. Going back to the
original expression $\tilde{p}_-(T_+)>p_+(T_+)$, and taking into
account the fact that $\tilde{p}_-$ is given by the
$\mathcal{O}(\phi^2)$ expansion of the thermal part of the effective
potential, one may use the approximation (\ref{vpot}) to write the
condition as $\epsilon>T_n^2 \phi_-^2\sum c_ig_ih_i^2/24$ or,
dividing by $a_+T_n^4$, as
\begin{equation}
\alpha_n> \alpha_{\mathrm{nec}}\equiv \frac{30}{\pi^2}
\left(\frac{\phi_-}{T_n}\right)^2
\frac{\sum c_ig_ih_i^2/24)}{\sum
c'_ig_i}. \label{alfainf}
\end{equation}
Of course, this condition is equivalent to Eq. (\ref{alfanec}), i.e.,
$\alpha_{\mathrm{nec}}={\alpha_c}/(1-\eta_\mathrm{UR}/ \epsilon)$.
Although Eq. (\ref{alfainf})  is discussed in Ref. \cite{ekns10}, the
simple phenomenological model used for explicit calculations
corresponds to setting $\eta_{\mathrm{UR}}=\eta_{\mathrm{NR}}$, which
gives  $\alpha_{\mathrm{nec}}={\alpha_c}/(1-\eta_\mathrm{NR}/
\epsilon)$. Similarly, in Ref. \cite{hs13} the dependence of the
microphysics on the wall velocity is discussed in some detail.
However, when hydrodynamics is included in the calculation, a
friction of the form $\eta_0/\gamma$ is assumed. As we have already
mentioned, this phenomenological model corresponds again to
$\eta_{\mathrm{UR}}=\eta_{\mathrm{NR}}$.

\subsection{Stationary and runaway regimes} \label{regimes}

\begin{figure}[bt]
\centering
\epsfysize=7cm \leavevmode \epsfbox{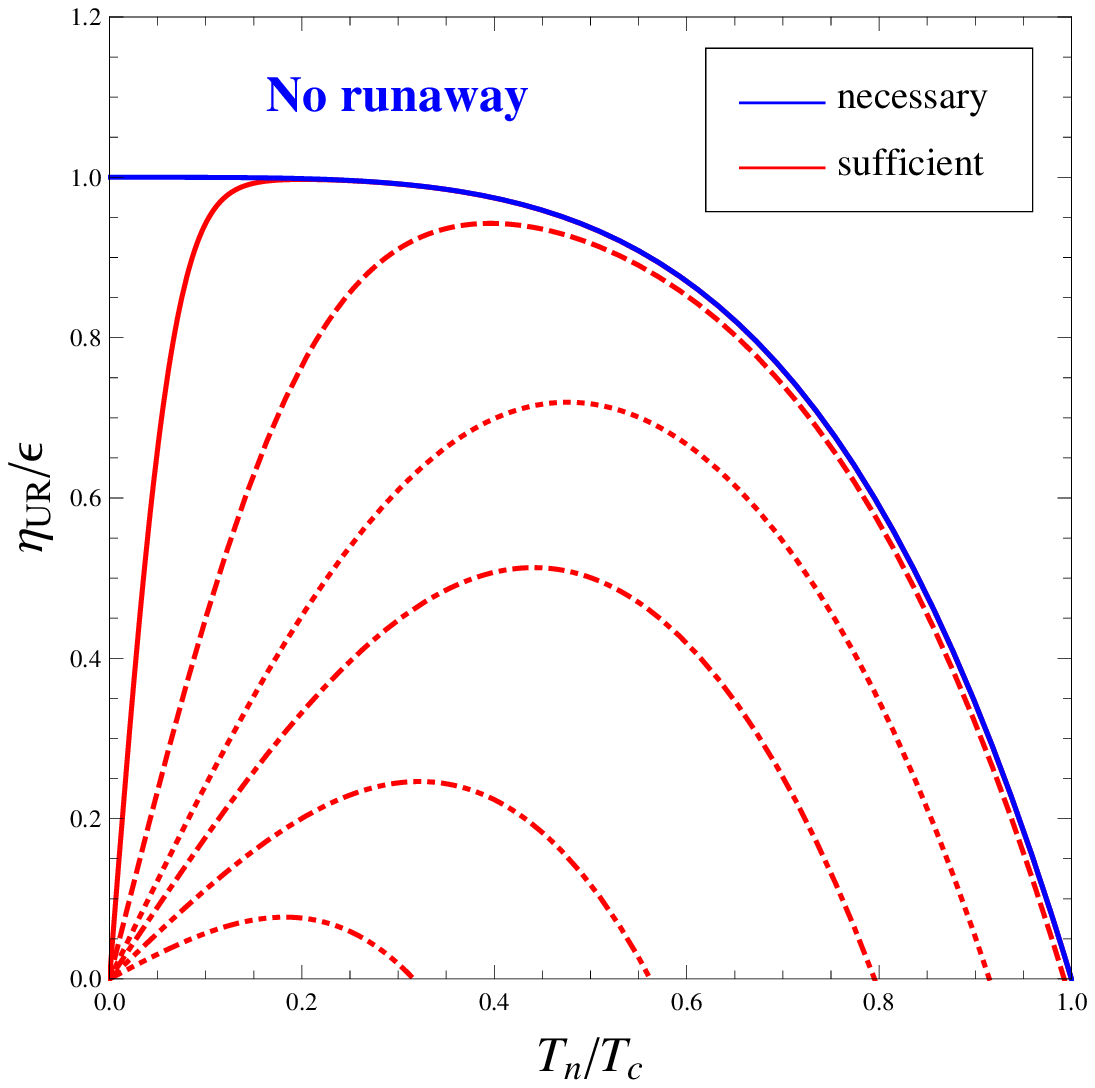}
\epsfysize=7cm \leavevmode \epsfbox{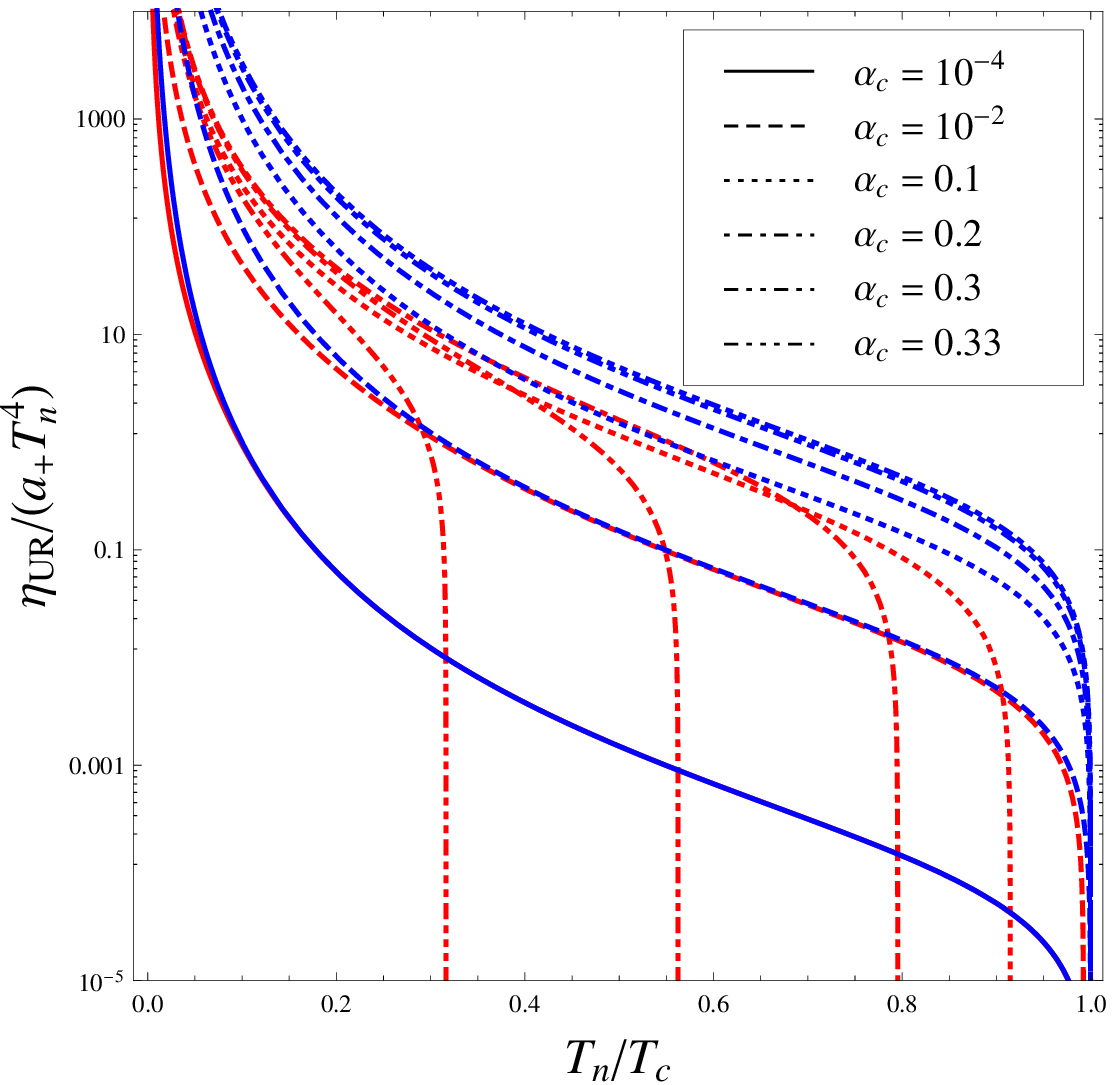}
\caption{Regions in the $(T_n,\eta_{\mathrm{UR}})$-plane where
runaway and detonation solutions can exist. The runaway necessary condition
(\ref{necessbag}) is fulfilled below the blue lines.
The sufficient  condition (\ref{sufibag})
is fulfilled below the red lines.}
\label{figcond}
\end{figure}

From Eq. (\ref{necessbag}) we see that the existence of runaway
solutions requires that both the UR friction parameter {and} the
nucleation temperature be small enough. In particular, the necessary
condition is never fulfilled for $\eta_{\mathrm{UR}}>\epsilon$, no
matter how small $T_n/T_c$. This is because the pressure difference
$\Delta p(T_n)$ is bounded by $\epsilon$ ($\Delta p\to \epsilon $ for
$T_n\to 0$) as much as the friction is bounded by
$\eta_{\mathrm{UR}}$. For $\eta_\mathrm{UR}$ smaller than, but close
to $\epsilon$, a strong supercooling is required [i.e., $(T_n/T_c)^4
\ll 1$]. Conversely, if the amount of supercooling is small (i.e.,
$T_n\approx T_c$), then ${\eta_\mathrm{UR}}\ll{\epsilon}$ is
required. The quantity $\epsilon=L/4$ is related to the strength of
the phase transition. Notice that the parameter
$\alpha_c=L/(4a_+T_c^4)$ is limited to the range $0<\alpha_c<1/3$.
The lower limit corresponds to a second-order phase transition with
$a_-=a_+$ and $L=0$. The higher limit corresponds to the case $a_-=0$
and $L=T_c s_+(T_c)$, i.e., to maximum entropy discontinuity at
$T=T_c$. Figure \ref{figcond} illustrates  the necessary condition
(\ref{necessbag}) and  the sufficient condition (\ref{sufibag}) in
parameter space.
In terms of $\eta_{\mathrm{UR}}/\epsilon$, the necessary condition
does not depend on the parameter $\alpha_c$ (left panel). In terms of
$\eta_{\mathrm{UR}}/(a_+ T_n^4)$, it does (right panel).

In the region above the blue lines the runaway solution does not
exist. Below these curves the runaway solution exists, but it may
coexist with stationary solutions. The red lines indicate the limits
of existence of detonation solutions. Thus, for a given value of
$\alpha_c$, the region above the corresponding red curve belongs to
detonations or other stationary solutions, whereas the runaway
solution is possible, in principle, only below the curve.
Qualitatively, the two conditions have a similar behavior in the
$(T_n,\eta_{\mathrm{UR}})$-plane. Both require small enough values of
$\eta_{\mathrm{UR}}$ and $T_n$. Quantitatively, the sufficient
condition is more restrictive and has a stronger dependence on
hydrodynamics. This is because,  for the stationary solution, the
released energy causes reheating and bulk motions of the fluid. The
runaway wall causes less perturbations, since the released energy
goes mainly into accelerating the wall.

Runaway solutions are driven by the pressure difference $\Delta
p(T_n)$. For $T_n$ close to $T_c$, we have $\Delta p(T_n)\approx 0$
and there won't be runaway solutions unless the friction parameter is
very small, as can be seen in Fig. \ref{figcond}. As we increase the
amount of supercooling, the pressure difference increases. For a
given value of $\eta_{\mathrm{UR}}/\epsilon$, one may in principle
expect that, as we keep increasing the amount of supercooling,
runaway solutions will eventually be possible. As we have seen,
though, the pressure difference saturates for strong supercooling (as
does the friction force for high wall velocity). Thus, the wall will
not run away for $\eta_{\mathrm{UR}}>\epsilon$ (see the left panel).
Below the blue curve, runaway solutions become possible. However,
whether the wall will run away or not, depends on the values of the
parameters. Consider for instance the case $\alpha_c=0.1$ (dotted
line) and a fixed value of $\eta_{\mathrm{UR}}$. If the value of
$\eta_{\mathrm{UR}}/\epsilon$ is, say, $0.8$, then the wall will
never run away, no matter how strong the supercooling. For a smaller
friction, say, $\eta_{\mathrm{UR}}/\epsilon=0.6$, then the wall will
run away for a strong enough supercooling ($T_n/T_c\lesssim 0.66$).
Eventually, for a very strong supercooling ($T_n/T_c\lesssim 0.29$)
we recover again the stationary solution.

This behavior is due to the added effects of friction and
hydrodynamics. Consider a weakly first-order phase transition, which
is characterized by a small amount of released energy (i.e., $L\ll
a_+T_c^4$, which implies $\alpha_c\ll 1$) and a small amount of
supercooling (i.e., $T_n\approx T_c$). In such a case, the effects of
hydrodynamics will disappear and the sufficient condition will
approach the necessary condition, as can be clearly seen in Fig.
\ref{figcond} (see, e.g., the cases $\alpha_c=10^{-4}$ and
$\alpha_c=10^{-2}$). A strong phase transition is characterized by a
large latent heat as well as a significant amount of supercooling.
For $T_n\ll T_c$ (and $\eta_{\mathrm{UR}}<\epsilon$), one expects
that the wall will run away. However, the release of latent heat
slows down the detonation, since the driving force is affected by
temperature gradients.  For a fixed value of the ratio $L/(a_+T_c^4)$
this effect becomes more and more important as the released energy
density $\Delta\rho\sim T_c^4$ becomes large\footnote{For the bag
model, the released energy density is bounded  below by
$\epsilon=L/4$.} in comparison with the plasma energy density
$\rho\sim T_n^4$. This is why the sufficient condition departs
further from the necessary condition for small $T_n$. The left panel
of figure \ref{figcond} shows that, for extremely supercooled phase
transitions, the friction must be very small for the wall to run
away.

In fact,  microphysics gives temperature-dependent friction
parameters. In a realistic model, we expect that $\eta_{\mathrm{UR}}$
will decrease at small temperatures, since the density of particles
in front of the wall vanishes for $T\to 0$. Hence, the right panel in
Fig. \ref{figcond} may be more illustrative for small temperatures.
Although the curves are more involved (since the necessary condition
now depends on $\alpha_c$) we see that, if we fix the value of
$\eta_{\mathrm{UR}}/(a_+ T_n^4)$, the system will always enter the
runaway region for strong enough supercooling. We shall discuss the
strong supercooling case (as well as the actual dependence of
friction on temperature) in Sec. \ref{strong}.

The behavior of the red curves can also be seen analytically. For
$\alpha_n\sim\alpha_c\ll 1$, Eq. (\ref{sufibag}) gives
$\eta_{\mathrm{suf}}/\epsilon=\eta_{\mathrm{nec}}/\epsilon-(3/8)
(\alpha_c/\alpha_n) (\alpha_n+\alpha_c)+\mathcal{O}(\alpha_c^2)$. The
negative sign of the $\mathcal{O}(\alpha)$ terms implies that the
value $\eta_{\mathrm{suf}}$ is smaller than $\eta_{\mathrm{nec}}$. As
we increase $\alpha_c$, the first term in Eq. (\ref{sufibag})
decreases whereas the second term increases. Hence, the value of
$\eta_{\mathrm{suf}}$ becomes smaller, departing from
$\eta_{\mathrm{nec}}$. As a consequence, the region where the wall
can runaway gets reduced. In the strong supercooling limit, we have a
linear behavior, $\eta_{\mathrm{suf}}/\epsilon
=(2/3^{1/4})(1/\alpha_c-3)^{1/4}(T_n/T_c) + \mathcal{O}(T_n/T_c)^3$.

\subsection{Stationary wall velocity}

\begin{figure}[hbp]
\centering
\epsfxsize=15cm \leavevmode \epsfbox{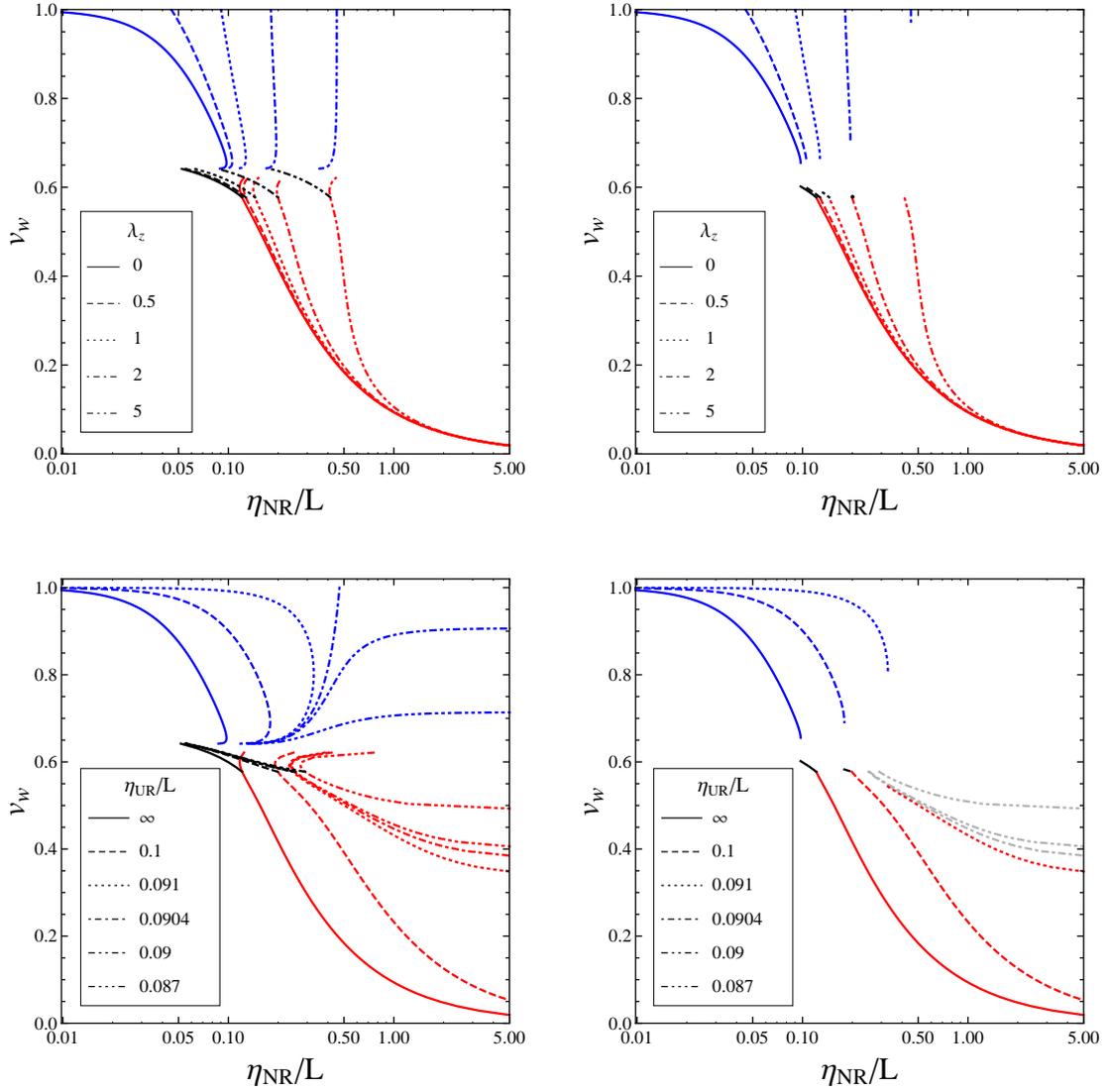}
\caption{The wall velocity
for the bag model with $\alpha_c=4.45\times 10^{-3}$ and $T_n=0.89T_c$.
Detonations are in blue, ``traditional'' deflagrations are in red,
and Jouguet deflagrations are in black.
The right panels show only the stable solutions.
}
\label{figvaretanr}
\end{figure}

We shall now study the stationary wall velocity as a function of the
parameters. We shall start the parameter variation from previously
considered values \cite{ikkl94,kl95,kl96,ms12}, which are convenient
to exhibit all the kinds of hydrodynamical solutions and to compare
with previous results for the old phenomenological friction model. In
Fig. \ref{figvaretanr} we considered the stationary wall velocity as
a function of the non-relativistic friction parameter
$\eta_{\mathrm{NR}}$.
Although we can vary independently $\eta_{\mathrm{NR}}$ and
$\eta_{\mathrm{UR}}$, in a real model there will be some correlation
between these parameters (e.g., both increase with the couplings of
particles to the Higgs). Therefore, we considered two opposite
relations for them. In the upper panels of Fig. \ref{figvaretanr} we
fixed the ratio $\lambda_z=\eta_{\mathrm{NR}}/\eta_{\mathrm{UR}}$,
whereas in the lower panels we fixed $\eta_{\mathrm{UR}}$. The left
panels show all the stationary solutions for a given set of
parameters. The right panel shows the solutions which are actually
realized in the phase transition.

Consider first the case of fixed $\lambda_z$ (upper panels). In the
left panel we see that, in some cases, there are several possible
solutions. In the first place, detonations and ``traditional''
deflagrations are bivalued in a certain range. This is not a problem,
since the lower branch of detonations and the upper branch of
deflagrations are unphysical solutions \cite{ms09} which, according
to numerical calculations, are unstable \cite{ikkl94,kl95,kl96} and,
thus, must be discarded\footnote{Notice that this eliminates the
Jouguet detonation (corresponding to the lower endpoint of the
detonation curves), which has often been considered for gravitational
wave generation.}. On the other hand, we see that also different
types of hydrodynamical solutions may coexist. Numerical calculations
seem to indicate that supersonic traditional deflagrations, which are
strong deflagrations, are unstable. Thus, the only supersonic
deflagration which can be realized is the Jouguet deflagration. In
case of coexistence of a deflagration with a detonation, the
detonation seems to be the stable solution \cite{ikkl94,kl95,kl96}.
We have plotted the curves in the right panel assuming this
hierarchy.

The case $\lambda_z=0$ (solid line) corresponds to the widely used
model (\ref{eqfieldnr}), which does not exhibit runaway behavior. For
$\lambda_z\neq 0$ the friction (\ref{fric}) saturates at high
velocities and the wall may run away. The runaway behavior depends
only on $\eta_{\mathrm{UR}}$. However, for fixed
$\lambda_z=\eta_{\mathrm{NR}}/\eta_{\mathrm{UR}}$, the wall always
runs away for small enough $\eta_{\mathrm{NR}}$, i.e., the wall
velocity becomes $v_w=1$ at
$\eta_{\mathrm{NR}}=\eta_{\mathrm{suf}}/\lambda_z$ (indicating that
the stationary wall assumption breaks down). Notice that the behavior
is qualitatively similar for the different values of $\lambda_z$.
Quantitatively, the differences become significant only for small
$\eta_{\mathrm{NR}}$ or large $\lambda_z$. The particular case
$\lambda_z=1$ corresponds to the models of Refs. \cite{ekns10,hs13},
for which the friction coefficient has the same value in the two
opposite regimes. Lower values of $\lambda_z$, corresponding to
higher values of the UR friction, yield a wider range of parameters
with stationary solutions.

In the lower panels of Fig. \ref{figvaretanr} we considered fixed
values of $\eta_{\mathrm{UR}}$. We have chosen most values  around
$\eta_{\mathrm{UR}}=\eta_{\mathrm{suf}}\simeq 0.0904$  (dashed-dotted
line).  For higher values of $\eta_{\mathrm{UR}}$ the curves
accumulate near the limiting line of $\eta_{\mathrm{UR}}\to\infty$
(which is the same as the curve of $\lambda_z=0$ in the upper
figure). For $\eta_{\mathrm{UR}}\leq\eta_{\mathrm{suf}}$  the
detonation solutions disappear (the curves in the left panel
correspond to the unphysical branches of the solutions, for which the
velocity increases with the friction), and the deflagrations approach
the speed of sound. The right panel shows only the physical
stationary solutions. It is interesting that for
$\eta_{\mathrm{UR}}\leq\eta_{\mathrm{suf}}$ there are still
deflagration solutions. We show these curves in grey. However, these
fast deflagrations coexist with the runaway solution, and it is
possible that they are unstable\footnote{In general, the traditional
deflagration seems to become unstable when it coexists with other
stationary solutions (namely, Jouguet deflagrations or detonations).
It is probable that the same happens when it coexists with the
runaway solution. A stability analysis is out of the scope of this
paper and shall be addressed elsewhere.}. Consequently, we shall
assume that for $\eta_{\mathrm{UR}}< \eta_{\mathrm{suf}}$ the wall
runs away. For $\eta_{\mathrm{UR}}> \eta_{\mathrm{suf}}$, we notice
that there is always a physical stationary solution, no matter how
small the value of the NR friction coefficient. This behavior
contrasts with the upper panel, where the wall always runs away for
small enough $\eta_{\mathrm{NR}}$. As we have already mentioned, in a
physical model there will always be a correlation between
$\eta_{\mathrm{NR}}$ and $\eta_{\mathrm{UR}}$. We do not expect, in
general, a very small $\eta_{\mathrm{NR}}$ together with a
considerably large $\eta_{\mathrm{UR}}$ or vice versa. As the
fundamental parameters of a model are varied, we expect an
intermediate behavior between the curves of the upper and lower
panels.

\begin{figure}[bt] \centering
\epsfxsize=7.5cm
\epsfbox{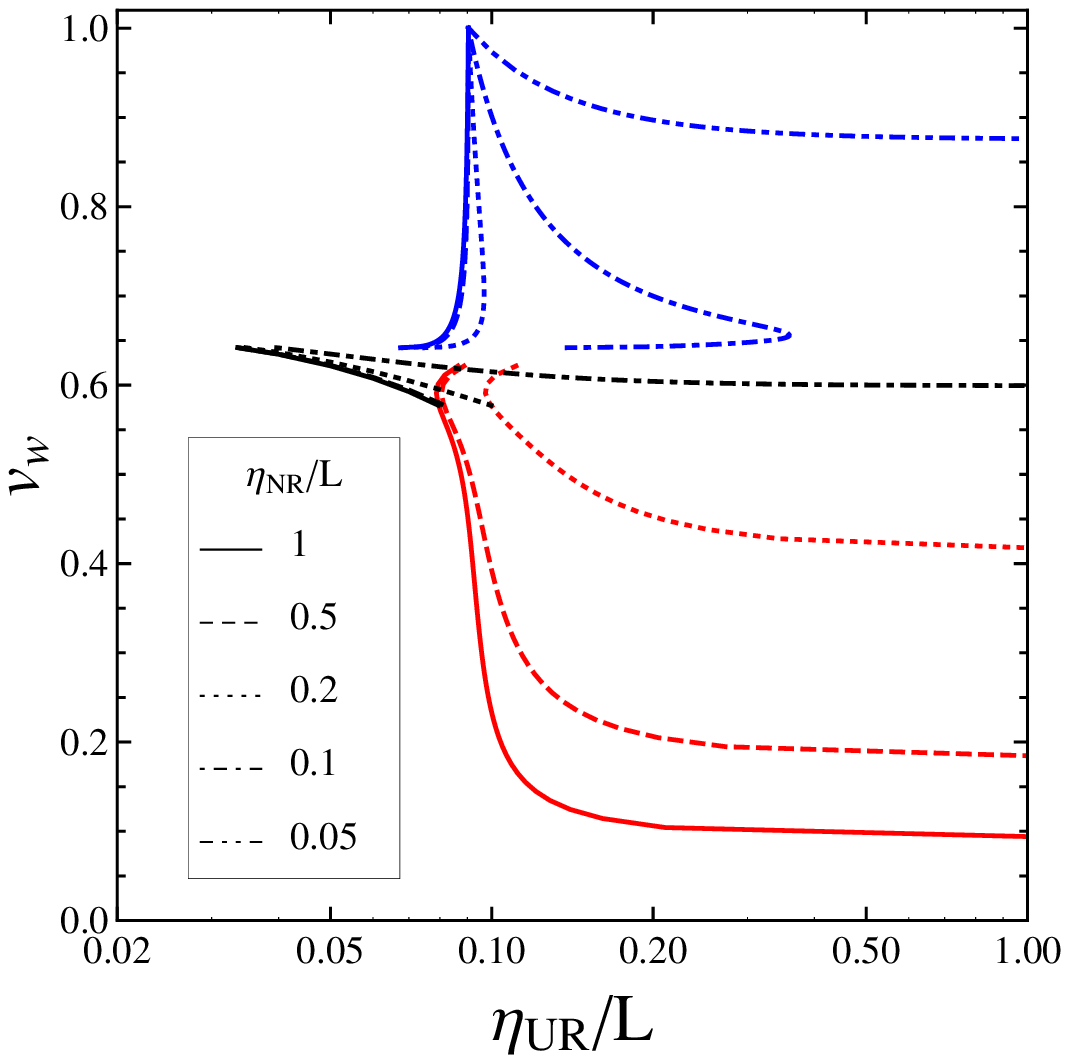}
\epsfxsize=7.5cm
\epsfbox{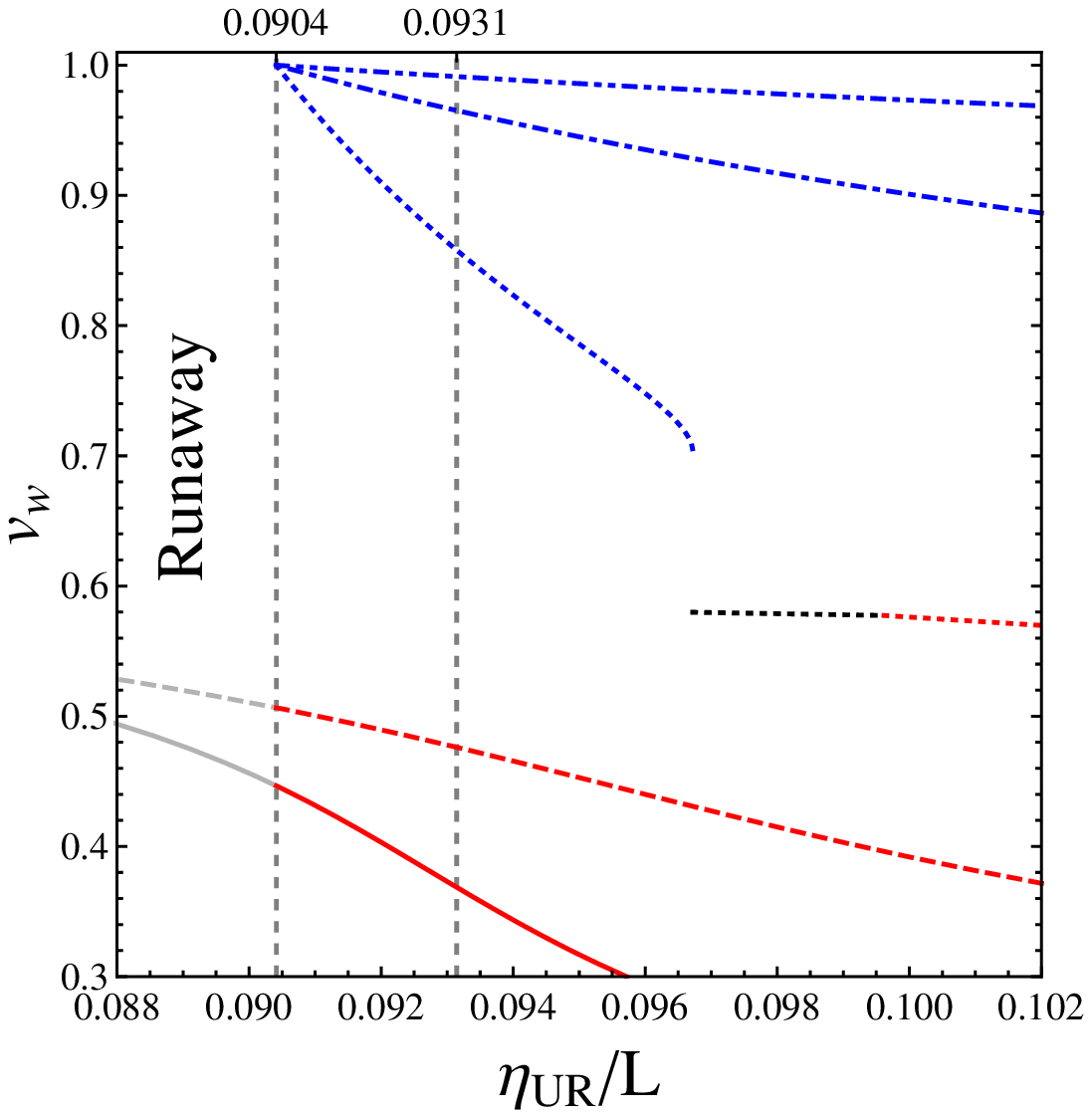}
\caption{The wall velocity as a
function of $\eta_{\mathrm{UR}}$, for the same bag parameters of Fig.
\ref{figvaretanr} and several values of $\eta_{\mathrm{NR}}$.
The vertical lines indicate the values of $\eta_{\mathrm{suf}}$
and $\eta_{\mathrm{nec}}$.}
\label{figetanrfijo}
\end{figure}

In Fig. \ref{figetanrfijo} we fixed $\eta_{\mathrm{NR}}$ and varied
$\eta_{\mathrm{UR}}$.
Notice that the wall velocity changes quickly near the critical value
$\eta_{\mathrm{UR}}=\eta_{\mathrm{suf}}$. For larger values the
velocity is essentially a constant which depends on the value of
$\eta_{\mathrm{NR}}$. As $\eta_{\mathrm{UR}}$ approaches
$\eta_{\mathrm{suf}}$ from the right, the velocity grows and the
detonation solutions disappear (their velocity becomes $v_w=1$).
Again, we see that for $\eta_{\mathrm{UR}}<\eta_{\mathrm{suf}}$ we
may still have physical deflagration solutions (in gray in the right
panel). As explained above, we shall assume that these are unstable
and we shall chose the runaway solution. In the right panel we have
zoomed the friction near $\eta_{\mathrm{UR}}=\eta_{\mathrm{suf}}$ and
we have marked the values of $\eta_{\mathrm{suf}}$ and
$\eta_{\mathrm{nec}}$  (which are very close to each other for the
present values of $\alpha_c$ and $\alpha_n$).

\begin{figure}[bt] \centering
\epsfxsize=7.5cm
\epsfbox{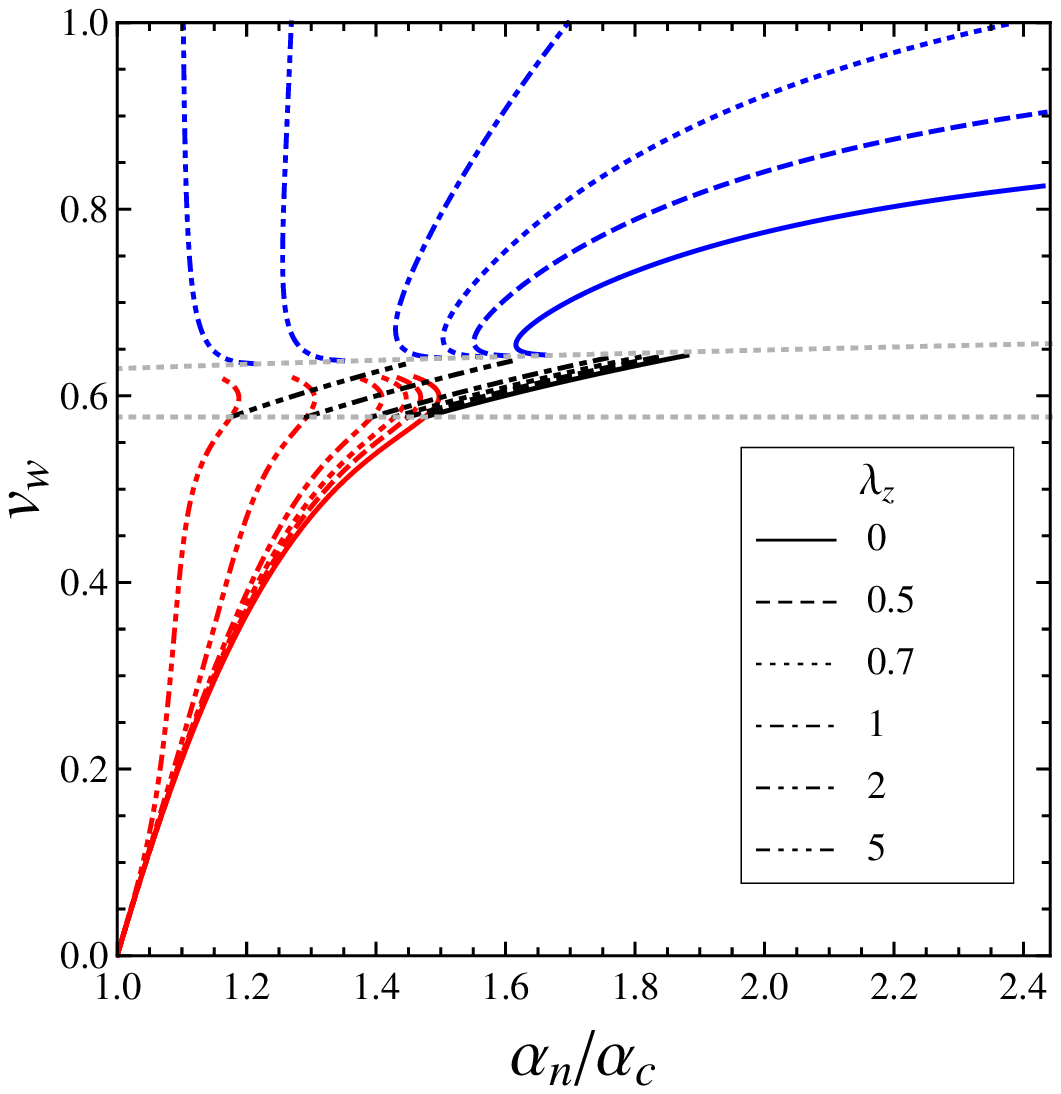}
\epsfxsize=7.5cm
\epsfbox{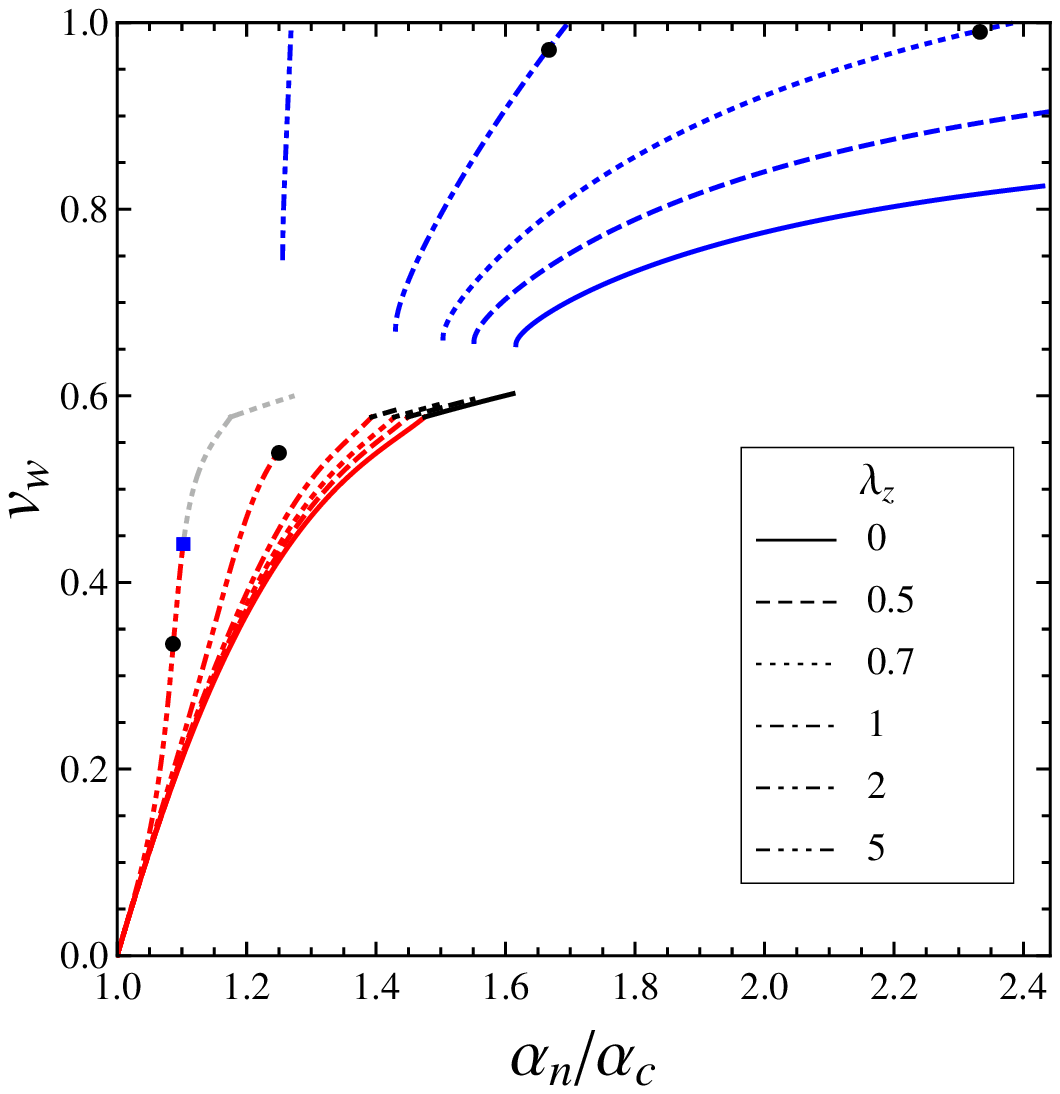}
\caption{The wall velocity as a
function of $\alpha_n/\alpha_c=(T_c/T_n)^4$, for $\alpha_c=4.45\times 10^{-3}$,
$\eta_{\mathrm{NR}}/L=0.1$, and several values of $\lambda_z$.}
\label{figvaralfan}
\end{figure}

Finally, in Fig. \ref{figvaralfan} we fixed the friction parameters
and varied the amount of supercooling.
In the left panel we also show the speed of sound and the Jouguet
detonation velocity $v_J^{\mathrm{det}}(\alpha_n)$ (dotted grey
lines).  We see that, for $T_n$ close to $T_c$, the solution is
always a weak deflagration and does not depend on the UR friction
parameter. The detonation solutions reach the speed of light at
$\alpha_n=\alpha_{\mathrm{suf}}$, indicating that beyond that value
the wall runs away. The runaway solutions exist already from
$\alpha_n=\alpha_{\mathrm{nec}}$, indicated in the right panel by
black round dots on the curves. For the smallest values of
$\lambda_z$ (highest values of $\eta_{\mathrm{UR}}$) the velocity
grows very slowly with the amount of supercooling. As we have
discussed earlier, the wall may eventually decrease for strong
supercooling, due to hydrodynamics effects. We shall analyze this
possibility in the next section. For high values of $\lambda_z$, the
ranges of stationary solutions get reduced (see the right panel).
This is also observed in the previous figures. For the case
$\lambda_z=2$, the Jouguet deflagration has disappeared (the weak
deflagration turns directly into a detonation) and the detonation has
a very short parameter range of existence. For $\lambda_z=5$ the
detonation has also disappeared and the weak deflagration turns
directly into a runaway solution at $\alpha_n=\alpha_{\mathrm{suf}}$
(we have marked this value with a squared blue dot, and we have
plotted the deflagration solution in grey beyond it). As can be seen
in the left panel, in this case it is the unphysical branch of
detonations that reaches the speed of light.

In all these figures, we see that the  detonation velocity becomes
$v_w=1$ at the values $\alpha_{\mathrm{suf}}$ or
$\eta_{\mathrm{suf}}$ obtained in sec. \ref{bag}. As we have just
seen, in some cases this actually corresponds to an unphysical
detonation. In general, cases like that of $\lambda_z=5$ in Fig.
\ref{figvaralfan}, where the physical detonation does not exist at
all, deflagrations do exist. Assuming, as we did, that in this case
the deflagration becomes unstable and jumps to a runaway solution at
$\alpha_n=\alpha_{\mathrm{suf}}$ (or $\eta_{\mathrm{UR}}
=\eta_{\mathrm{suf}}$ in the right panel of Fig. \ref{figetanrfijo}),
the values of $\alpha_{\mathrm{suf}}$ and $\eta_{\mathrm{suf}}$ still
make sense as limits between a stationary solution (either a
detonation or a deflagration) to a runaway solution. This is the
actual meaning of the red curves in Fig. \ref{figcond}.

\section{Strong supercooling} \label{strong}

The fast propagating modes of a phase transition front, namely,
detonations and runaway walls, are interesting due to the possibility
of generating sizeable gravitational waves. The high velocity
behavior is governed by the UR friction parameter. As we have seen,
though, whether the wall will be  fast or not, depends also on the
amount of supercooling  and on the latent heat. A considerable amount
of supercooling favors large pressure differences and high
velocities, whereas a large release of latent heat slows down the
wall.  Strong supercooling is typical of strongly first-order phase
transitions. However, the latter are also characterized by a large
latent heat.  On the other hand, at low temperatures one expects a
small friction.

The friction depends strongly on the  values of the particle masses
inside the bubble (assuming for simplicity that $m_i=0$ outside);
more precisely, on the ratio $m_i/T$. Thus, a relevant parameter is
the value of $\phi/T$ in the broken-symmetry phase. Strongly
first-order phase transitions are characterized by a relatively high
value of the ratio $\phi_0(T_c)/T_c$. This implies also a relatively
high latent heat. Nevertheless, $\phi_0$ and $T_c$ are both given by
the characteristic scale of the theory, and its ratio will be
$\mathcal{O}(1)$ for a natural phase transition. As we have seen, for
the bag EOS, $L$ is bounded by $\frac{4}{3}a_+T_c^4$. On the other
hand, in some models the barrier between minima may persist at low
temperatures, causing a large amount of supercooling. In such a case
we may have $T_n\ll T_c$, which implies $\phi_0(T_n)/T_n\gg 1$ and
$L\gg a_+T_n^4$.

On the whole,  for very strong supercooling we will have in principle
a high pressure difference between phases and a small friction force,
but the energy released will be large in comparison to the energy
density of the plasma, and hydrodynamics effects will be important.
It is thus worth investigating the behavior of phase transition
fronts in such a case.

\subsection{Detonations and runaway solutions}

\begin{figure}[tb] \centering
\epsfysize=7cm
\epsfbox{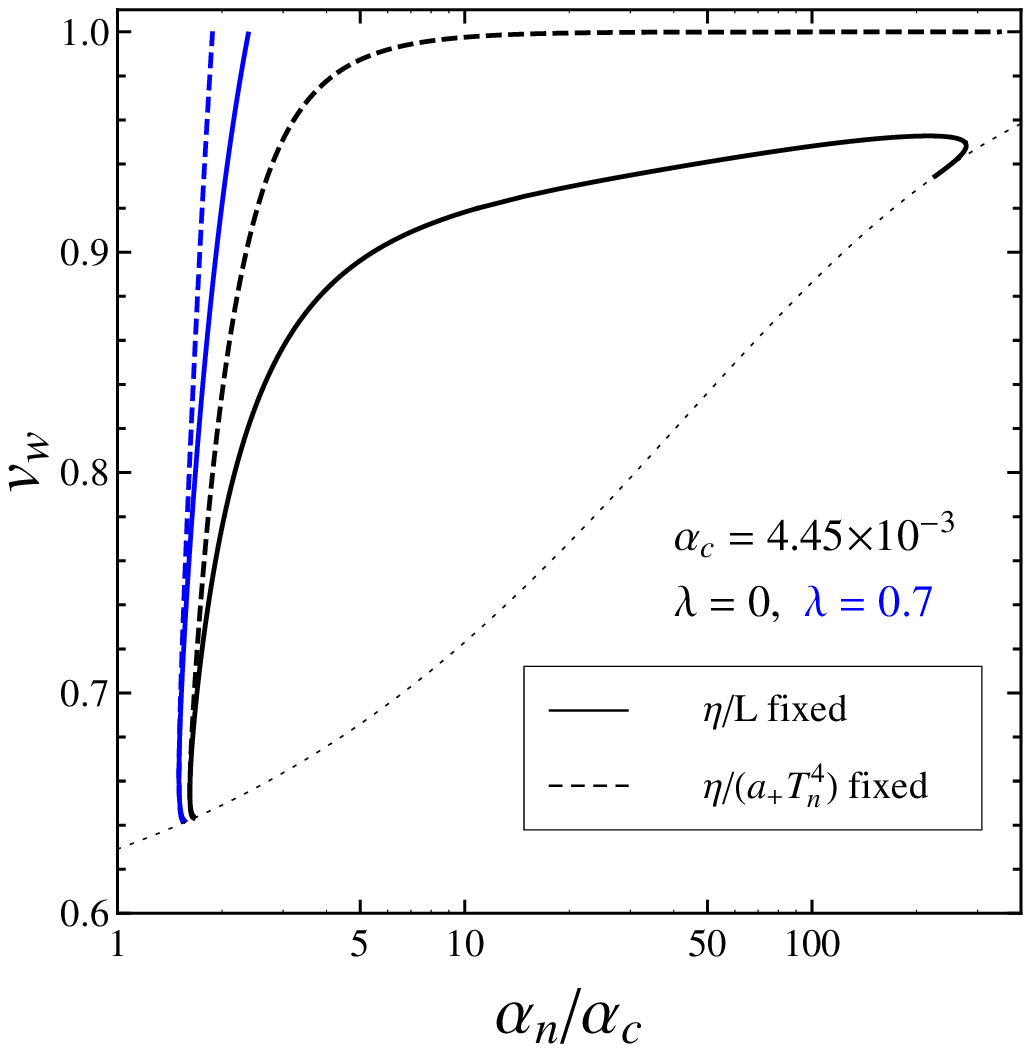}
\epsfysize=7cm
\epsfbox{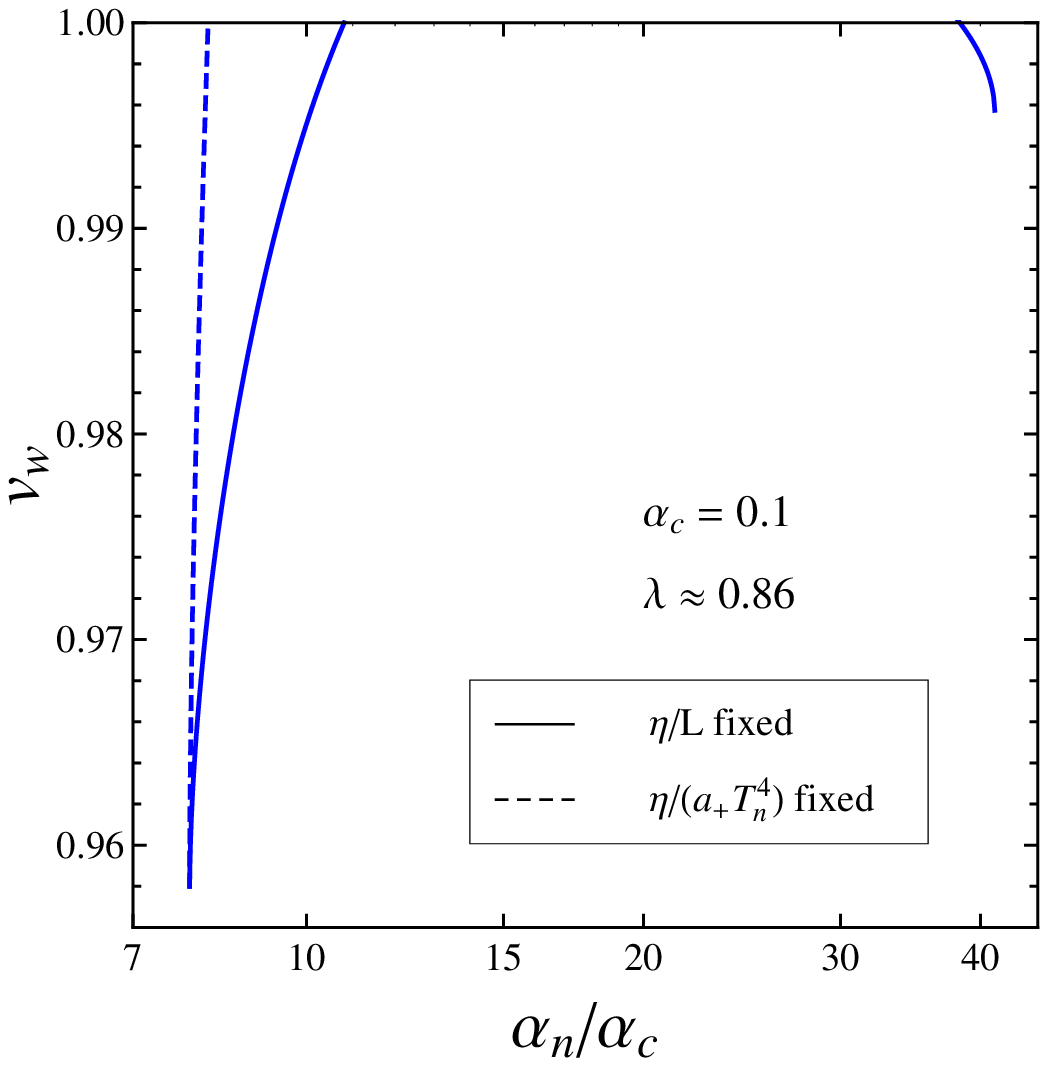}
\caption{The wall velocity as a
function of $\alpha_n/\alpha_c$ for $\eta/L$ fixed (solid)
and $\eta/(a_+ T_n^4)$ fixed (dashed). In the left panel we have
$\eta_{\mathrm{NR}}=0.1 L$ for $T_n$ close to $T_c$. In the right panel we have
$\eta_{\mathrm{NR}}=0.15 L$ and $\eta_{\mathrm{UR}}=0.175 L$
for $T_n$ close to $T_c$.
}
\label{figsuperc}
\end{figure}

As we have seen, strong hydrodynamics effects may prevent the wall
from reaching ultra-relativistic velocities, even in the case of a
strong supercooling.  To illustrate this effect, let us plot again
$v_w$ vs. $\alpha_n$ like in Fig. \ref{figvaralfan}, this time for a
wider range of the supercooling parameter $\alpha_n/\alpha_c$. It is
interesting to consider first the same thermodynamic parameters of
Fig. \ref{figvaralfan}, which correspond to a relatively low value of
the released energy (since $L\sim 10^{-2}a_+T_c^4$). In the left
panel of Fig. \ref{figsuperc} we consider a couple of representative
values of $\lambda_z$.
As an example of the case $\eta_{\mathrm{UR}}>\epsilon$, for which
the wall will not run away, we choose $\lambda_z=0$, i.e., the limit
$\eta_{\mathrm{UR}}\to \infty$ (black lines). We see that, for fixed
values of the friction parameters (solid black line), the velocity
grows slowly with the amount of supercooling, and eventually
decreases for very strong supercooling. Moreover, for
$\alpha_n/\alpha_c\gtrsim 300$ the stationary solution cannot exist
due to strong hydrodynamics effects. Before disappearing, the
detonation approaches the Jouguet point (dotted grey line) and a
non-physical branch of solutions appears, just like at the other end
of the curve (where the supercooling becomes insufficient for a
detonation solution). As we have mentioned, for strong supercooling
it is more realistic\footnote{We discuss this issue below.} to
consider a friction which decreases as $T_n$ decreases. Therefore, we
have also considered fixed values of $\eta/(a_+ T_n^4)$ (dashed black
line). In this case, the wall velocity is only bounded by the
relativistic limit $v_w=1$. The wall does not runaway because we are
considering the case $\eta_{\mathrm{UR}}\to \infty$. We also consider
the case $\lambda_z=0.7$ (blue lines). The solid line ($\eta/L$
fixed) corresponds to the dotted line in Fig. \ref{figvaralfan},
which runs away for $\alpha_n/\alpha_c\simeq 2.4$. For $\eta/(a_+
T_n^4)$ fixed (dashed blue line) the wall runs away at a smaller
$\alpha_n$.

Hydrodynamics effects will be stronger for higher values of the
latent heat. In Sec. \ref{regimes} we have discussed the case
$\alpha_c=0.1$, corresponding to the dotted line in Fig.
\ref{figcond}. According to that figure (see the left panel), for
$\eta_{\mathrm{UR}}/L\approx 0.15$ (i.e.,
$\eta_{\mathrm{UR}}/\epsilon\approx 0.6$) the detonation solution
gives way to the runaway solution at a given amount of supercooling.
But then, for a stronger amount of supercooling, the detonation
solution is possible again. The right panel of Fig. \ref{figsuperc}
shows this effect (we have only kept the physical solutions and we
have chosen the values $\eta_{\mathrm{NR}}/L= 0.15$ and
$\eta_{\mathrm{UR}}/L= 0.175$, which better illustrate the effect).
It is important to notice, however, that shortly after reappearing,
the detonation ceases to exist due to strong hydrodynamics. In many
cases, the detonation will not reappear at all. In any case, if we
consider a friction which vanishes at zero temperature (dashed line),
then a smaller supercooling will suffice for the wall to run away,
and the stationary solution will not appear again at stronger
supercooling. This can be seen already in the right panel of Fig.
\ref{figcond}.

So far we have considered two simple cases for the variation of the friction
coefficients with temperature, which consist of fixing the dimensionless
ratios $\eta/L$ and $\eta/(a_+T_n^4)$. In a realistic model, we expect
temperature-dependent friction parameters, although not necessarily
decreasing as fast as  $T^4$. Roughly, the number density of massless
particles in the symmetric phase goes as $T^3$, and the force each particle
feels at the wall is proportional to $\Delta p\approx (m/p)\Delta m$ (in the
WKB approximation). For $\Delta m=m(\phi_0)$, we have $\Delta p\sim m^2/T$.
Thus we obtain a force of the form $F\sim m^2 T^2$. As we shall see, this is
parametrically correct for $\eta_{\mathrm{UR}}$. The case of
$\eta_{\mathrm{NR}}$ is more complex because one must take into account
particle interactions. For interaction rates $\Gamma\sim T$, one obtains
again $F_{\mathrm{fr}}\propto T^2$ (see below).

\begin{figure}[htb] \centering
\epsfysize=7cm
\epsfbox{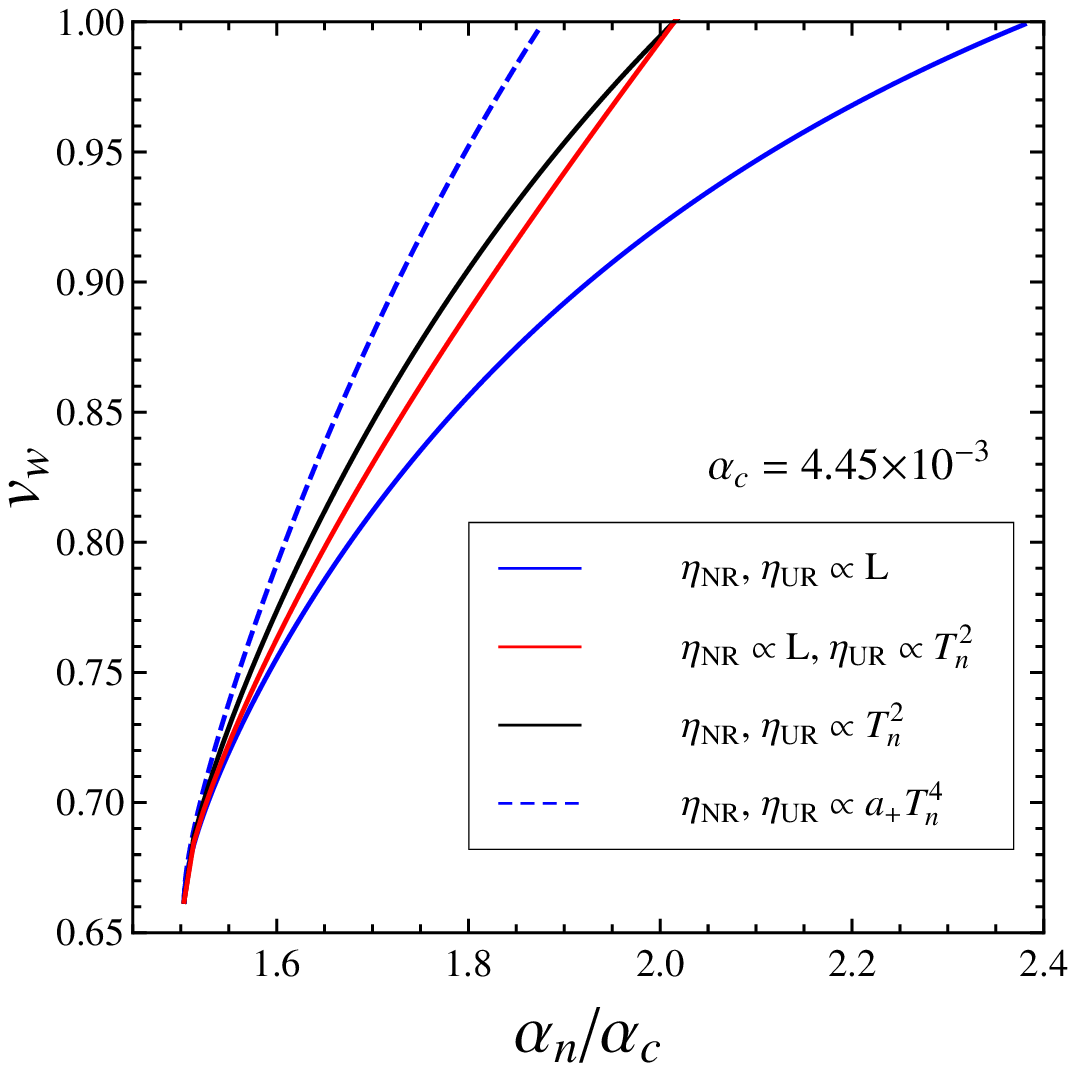}
\epsfysize=7cm
\epsfbox{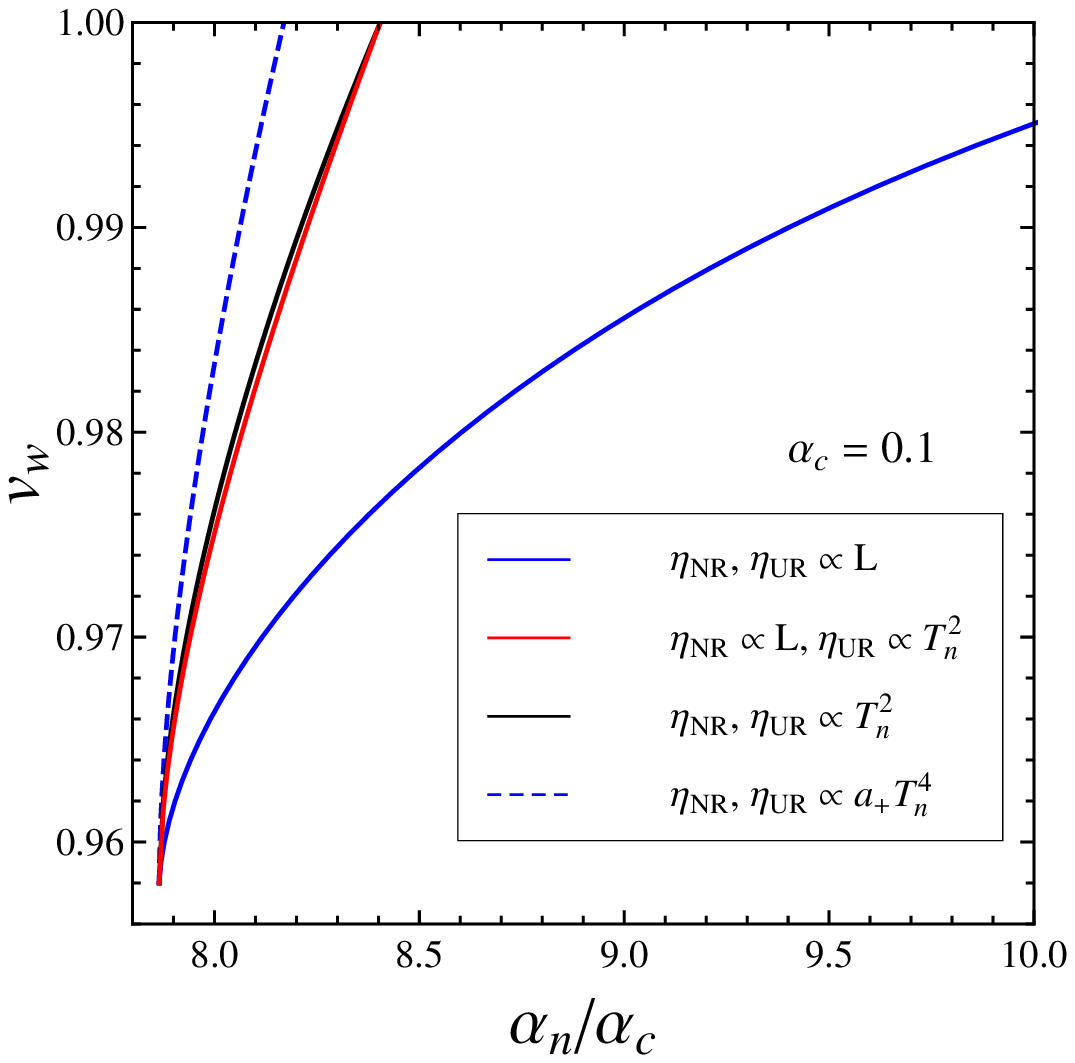}
\caption{The same as Fig. \ref{figsuperc} (blue lines), but we have included the
cases $\eta_{\mathrm{NR}},\eta_{\mathrm{UR}}\propto T_n^2$ (black lines) and
$\eta_{\mathrm{UR}}\propto T_n^2$, $\eta_{\mathrm{NR}}=$ constant (red lines).
}
\label{figfrict}
\end{figure}

To investigate the behavior of the wall velocity with a friction
which depends quadratically on temperature, we shall consider
friction parameters of the form $\eta=\eta_0
T_n^2=\eta_0'\sqrt{\alpha_c/\alpha_n}$. We choose the coefficient
$\eta_0$ so that the different forms of the friction match for
$T_n\sim T_c$. In order to evaluate to what extent fast solutions
depend on the NR friction parameter, we shall also consider the case
of a constant $\eta_{\mathrm{NR}}$  with a quadratic
$\eta_{\mathrm{UR}}$. We show the result in Fig. \ref{figfrict},
together with the curves of Fig. \ref{figsuperc} for comparison. The
black lines correspond to both friction coefficients of the form
$\eta=\eta_0 T^2$. The red lines correspond to a constant NR friction
coefficient.
We see that  the black and red curves are very close to each other,
and meet in the limit $v_w= 1$. This is because the value of the NR
friction coefficient does not play a relevant role for detonations,
and becomes irrelevant in the ultra-relativistic limit. In the right
panel the effects of hydrodynamics are stronger since the released
energy is larger (notice the larger amounts of supercooling needed to
obtain detonations and runaway walls). We see that the solid line
deviates significantly from the others. This corresponds to the
unrealistic case in which the UR friction is temperature-independent.

\subsection{Strong supercooling and friction coefficients}

In this last subsection we wish to discuss briefly the behavior of
the friction coefficients at low nucleation temperatures (see
\cite{ms10} for other limiting cases). A detailed analysis is out of
the scope of the present paper. Therefore, we shall assume that the
WKB approximation is still valid. For very low temperatures, such
that $T\ll l_w^{-1}$ this approximation will certainly break down.

Let us first consider the non-relativistic friction coefficient
$\eta_{\mathrm{NR}}$. Since the interaction rates are proportional to
temperature, at low temperatures the approximation $\Gamma\gg
l_w^{-1}$ which leads to Eq. (\ref{eta}) is no longer valid.
Nevertheless, we shall consider the approximation (\ref{eta}) to
catch a glimpse of the behavior of the friction. In any case, we have
just seen that the wall velocity depends very weakly on
$\eta_{\mathrm{NR}}$ for strong supercooling. In the limit $m/T\gg 1$
the coefficient $c_1$ is suppressed by a Boltzmann factor,
\begin{equation}
c_{1}\approx\left( m/T\right) ^{1/2}\exp \left( -m/T\right) /\left( 2\pi \right)
^{3/2}.  \label{c1gde}
\end{equation}%
However, in the case of interest, very light particles in the
symmetric phase become heavy in the broken-symmetry phase. For
simplicity, let us consider masses of the form $m_i(\phi)=h_i\phi$.
To calculate the integral in Eq. (\ref{eta}), we notice that for
small $\phi$ the integrand is suppressed by powers of $\phi$.
Therefore we can use the approximation (\ref{c1gde}) in all the range
of integration. We have
\begin{equation}
\eta_\mathrm{NR} =\sum_{i}\frac{g_{i}T^4}{\Gamma (2\pi)^3}
\int_{0 }^{h\phi_0/T}
e^{-2x}x ^{3}\left|\frac{dx}{dz}\right|dx, \label{etanrlowt}
\end{equation}%
with $x=h_i\phi/T$. To perform the integral, an approximation for
$\phi(z)$ is needed. Instead of the usual $(\phi_0/2)([1 + \tanh
(z/l_w)]$, we shall use the much simpler approximation of a linear
function $\phi=\phi_0/l_w\, z$ inside the wall and $\phi=$ constant
outside the wall. The temperature dependence, as well as the
parametric behavior, should not depend on the approximation for the
wall profile. Thus, in the integral (\ref{etanrlowt}) $dx/dz$ is a
constant and we obtain, for $h\phi_0/T\gg 1$,
\begin{equation}
\eta_\mathrm{NR} =\sum_{i}
\frac{3g_{i}h_i\phi_0 T^2}{64\pi^3(\Gamma/T) l_w}.
\end{equation}%
We see that the friction coefficient vanishes for small $T$, though not as
fast as $T^4$.

The ultra-relativistic friction coefficient is very easy to estimate
in the strong supercooling limit, either from its definition in Eqs
(\ref{fricpmen}-\ref{fricurmicro}) or directly from the total force
(\ref{totalforce0}). It is interesting to consider the latter. For
$\phi_+=0$ and $m_i=h_i\phi$, the total force is given, \emph{at any
temperature}, by
\begin{equation}
\frac{F}{A}=V(\phi_+)-V(\phi_-)-\sum_i g_i c_i  h_i^2\frac{\phi_o^2
T^2}{24}. \label{totftchica}
\end{equation}
At high temperatures, the last term in Eq. (\ref{totftchica}) is part
of the equilibrium pressure difference [as can be seen in the
high-temperature expansion (\ref{vpot})]. Hence, it is part of the
driving force. As a consequence, the friction is given by the rest of
the high-$T$ expansion, $\eta_{\mathrm{UR}}\sim
T\phi_0^3+\mathcal{O}(\phi_0^4)$ [see Eq. (\ref{fricur})]. At very
low temperatures, on the contrary, the driving force is just given by
the vacuum potential, i.e., the first two terms in Eq.
(\ref{totftchica}). Hence, the last term now gives the friction
force, and we have
\begin{equation}
\eta_{\mathrm{UR}}=\sum_i g_i c_i  h_i^2\frac{\phi_o^2
T^2}{24}.
\end{equation}
We see that $\eta_{\mathrm{UR}}$ goes as $T^2$ at low temperatures.

\section{Conclusions} \label{conclu}

In this paper we have studied the friction force acting on phase
transition fronts. In particular, we have discussed the
ultra-relativistic behavior of the friction, derived in Ref.
\cite{bm09}. We have considered runaway walls as well as stationary
walls (detonations), which have different hydrodynamics. We have
shown that the two solutions coexist in a range of parameters. Thus,
we have argued that the runaway condition given in Ref. \cite{bm09}
is only a necessary condition and does not guarantee that the wall
will run away.  We have found a sufficient condition for the wall to
run away by using the criterion of the detonation reaching the speed
of light.

We have also proposed a phenomenological model for the friction,
which interpolates between the non-relativistic and the
ultra-relativistic behaviors. The main improvement of this model with
respect to previous ones is the incorporation of a new free parameter
governing the ultra-relativistic limit of the friction force. The
value of this parameter can be easily calculated for any specific
model. Therefore, our model not only gives a saturating friction, but
also gives the correct quantitative behavior of the friction in the
UR limit. This phenomenological model is consistent with the
necessary and sufficient runaway conditions, which do not depend on
the non-relativistic friction coefficient.

Using the bag equation of state and our phenomenological model for
the friction, we have studied the runaway conditions, as well as the
stationary regime, as functions of the thermodynamic and friction
parameters. Regarding the runaway conditions, the general result is
that they are quantitatively similar for small values of the latent
heat but depart otherwise. If we, for instance, increase the amount
of supercooling (for a given value of the friction), then, in the
case of small latent heat, the sufficient condition is reached
shortly after the necessary condition. On the contrary, for larger
latent heat, the larger release of energy slows down the phase
transition fronts. This causes the detonation to persist for larger
amounts of supercooling, even if the runaway solution already exists.
Regarding the stationary wall velocity, we have found, as expected,
that for parameters which favor a small wall velocity (i.e., for low
supercooling, large latent heat, or high non-relativistic friction)
the solution depends very weakly on the UR friction coefficient.
Conversely, for parameters which favor fast stationary solutions, the
velocity does not depend significantly on the NR friction
coefficient.

We have studied in particular the case of a phase transition with a
large amount of supercooling, which is important for the generation
of gravitational waves. In this case the dependence of the friction
on temperature becomes relevant. We have explored the wall velocity
for different behaviors of the friction coefficients. Thus, we have
considered fixed values of the dimensionless parameters $\eta/L$ and
$\eta/T^4$, as well as the quadratic dependence $\eta\propto T^2$.
The latter is motivated by low-temperature approximations, which we
have also discussed. The value of the NR friction parameter  does not
affect the strong supercooling case, since the velocity is in general
high. The general result is that, if the UR friction parameter does
not decrease for strong supercooling, then the wall may reach a
saturation value, and even decrease due to hydrodynamics effects. In
contrast, if the UR friction parameter vanishes at zero temperature
(which seems to be the actual case), then the wall always runs away
for strong enough supercooling.

\section*{Acknowledgements}

This work was supported by Universidad Nacional de Mar del Plata,
Argentina, grant EXA 607/12.

\end{document}